\documentclass[preprint]{aastex701}

\usepackage{hyperref}
\accepted{December 15, 2025}
\journalinfo{}

\begin{document}

\title{Tracking the Evolution of Plasma Instabilities from the Prominence-Corona Transition Region into Interplanetary Space with Total Solar Eclipse and WISPR/PSP White Light Images}

\correspondingauthor{Shadia R. Habbal}

\author[orcid=0000-0003-4089-9316,gname=Shadia R., sname=Habbal]{Shadia R. Habbal} 
\affiliation{Institute for Astronomy, University of Hawai`i, 2680 Woodlawn Drive, Honolulu, HI 96822, USA}
\email[show]{habbal@hawaii.edu}

\author[orcid=0000-0002-3089-3431,gname=Shaheda Begum, sname=Shaik]{Shaheda Begum Shaik}
\affiliation{George Mason University, Fairfax, VA 22030, USA}
\altaffiliation{U.S. Naval Research Laboratory, Washington, DC 20375, USA}
\email{sshaik7@gmu.edu}

\author[orcid=0009-0002-9883-278X,gname=Zachary, sname=Bailey]{Zachary Bailey}
\affiliation{Institute for Astronomy, University of Hawai`i, 2680 Woodlawn Drive, Honolulu, HI 96822, USA}
\email{zbailey@hawaii.edu}

\author[orcid=0000-0001-5207-9628,gname=Nathalia, sname=Alzate]{Nathalia Alzate}
\affiliation{ADNET Systems, Inc., Greenbelt, MD 20771, USA}
\altaffiliation{NASA Goddard Space Flight Center, Greenbelt, MD 20771, USA}
\email{nathalia.alzate@nasa.gov}

\author[orcid=0000-0002-6962-0959,gname=Riddhi, sname=Bandyopadhyay]{Riddhi Bandyopadhyay}
\affiliation{Department of Astrophysical Sciences, Princeton University, Princeton, NJ 08544, USA}
\email{riddhib@princeton.edu}

\author[orcid=0000-0001-7312-2410,gname=Miloslav, sname=Druckm{\"u}ller]{Miloslav Druckm{\"u}ller}
\affiliation{Faculty of Mechanical Engineering, Brno University of Technology; Brno, 616 69, Czech Republic}
\email{druckmuller@fme.vutbr.cz}

\author[orcid=0000-0001-6407-7574,gname=Simone, sname=Di Matteo]{Simone Di Matteo}
\affiliation{The Catholic University of America, Washington, DC,20064, USA}
\altaffiliation{NASA Goddard Space Flight Center, Greenbelt, MD 20771, USA}
\email{simone.dimatteo@nasa.gov}

\author[orcid=0000-0002-6396-8209,gname=Sage, sname=Constantinou]{Sage Constantinou}
\affiliation{Institute for Astronomy, University of Hawai`i, 2680 Woodlawn Drive, Honolulu, HI 96822, USA}
\email{sagecons@hawaii.edu}

\begin{abstract}
High-resolution total solar eclipse (TSE) white light (WL) images are the only observations at present to capture coronal structures over an uninterrupted field of view (FoV) of at least 10 solar radii ($R_s$) starting from the solar limb. They were the first to report the presence of vortex rings originating within the prominence-corona transition region (PCTR). They also captured CMEs and Kelvin-Helmholtz (KH) instabilities at different phases of their evolution. While the evolution of CMEs and KH waves is relatively well-documented, little is known about the survivability of vortex rings beyond the FoV of the TSE images. In this study, we use seven TSE images and non-contemporaneous WL images acquired by the Wide-Field Imager for Parker Solar PRobe (WISPR) to track the spatial evolution of vortex rings, KH waves, and CMEs. The size trend versus radial distance for vortex rings and KH waves are found to be shallower below 1.5 $R_s$ than beyond 3 $R_s$, while the CMEs observed beyond 3 $R_s$ show a unique slope. The WISPR time series yield an average speed of 249.02 $\pm$ 25.3 km/s for the vortex rings beyond 3 $R_s$, that when combined with their size yields a speed of 19.39 $\pm$ 3.20 km/s below 1 $R_s$. These values are remarkably consistent with the acceleration profile of the slow solar wind over the same distance. This study provides strong empirical evidence that vortex rings, which originate at the PCTR with complex magnetic structures, do not dissipate as they expand away from the Sun with the solar wind.
\end{abstract}


\section{Introduction} \label{sec:intro}

Total solar eclipses (TSEs) were the first to reveal the presence of pinkish filamentary features, morphologically distinct from the coronal structures through which they protruded. Referred to as prominences, the first spectroscopic TSE observations attributed their pinkish hue to  H$\alpha$ emission characteristic of chromospheric plasmas at around $10^4$ K  \citep{janssen1869a,janssen1869b}. 
\cite{Ohman_1969} was probably the first to provide a comprehensive account of the rotational/helical motions in prominences based on reports by \cite{Pettit_1925}, as well as from his own observations of smoke-ring type features in the vicinity of a sunspot \citep{Ohman_1968}. \cite{Liggett_Zirin_1984} later showed that approximately 10\% of non-eruptive limb prominences exhibit localized or general rotational motions, on scales of a few to 10’s of Mm. Taking advantage of adaptive optics to reach the diffraction limit of the 1.6-m Goode Solar Telescope (GST),  \citet{Schmidt_2025} recently presented novel manifestations of prominence dynamics, which they referred to as `rain', down to the 0.01 Mm spatial resolution of the telescope.

\begin{figure}[!h]
 \centering
 \includegraphics [width=0.65\textwidth]{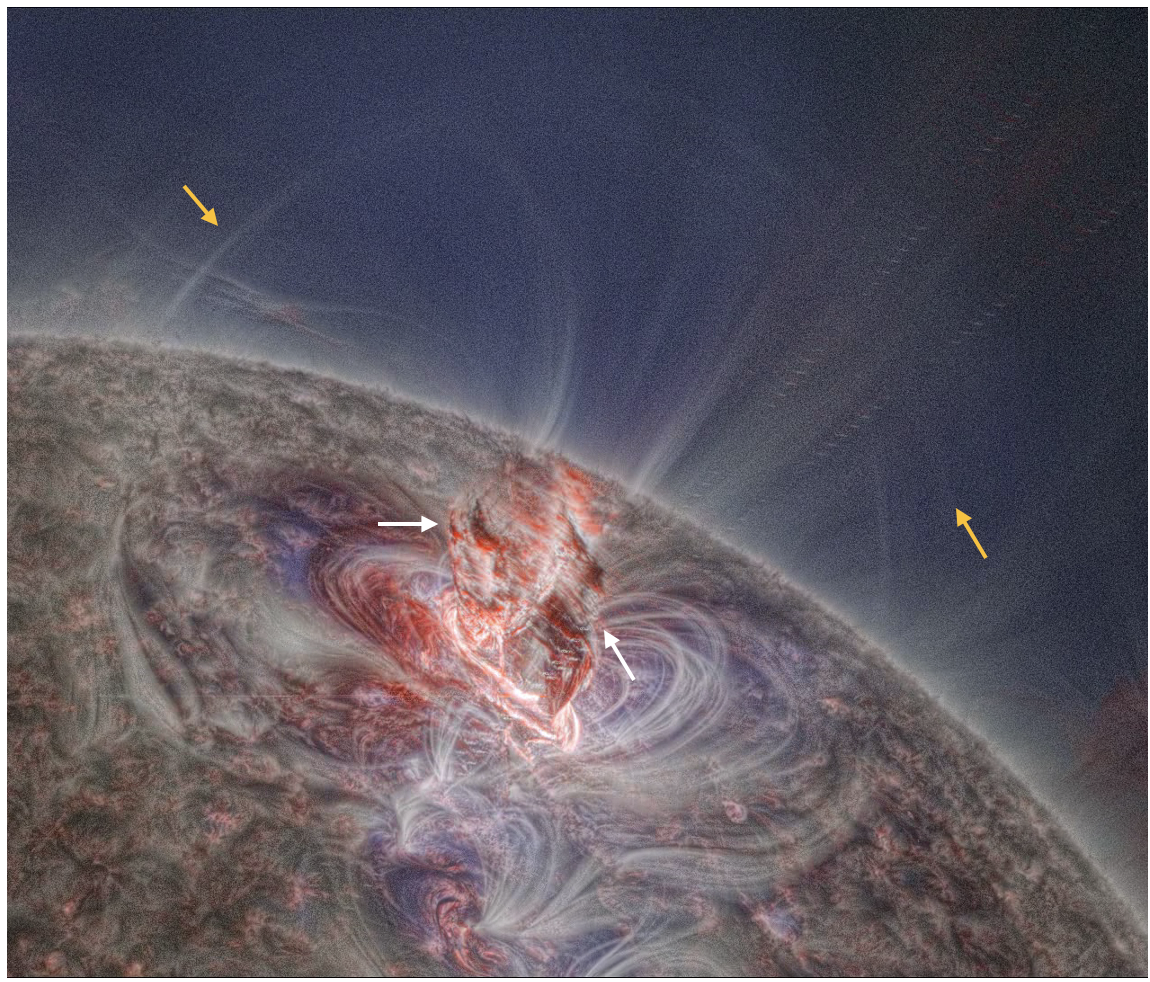} 
\caption{\label{SDOarrows} 
{\footnotesize Prominence eruption during the flare of 2011 June 7 as captured by SDO/AIA as composite images from the 304, 171, and 211 \AA\ channels in SDO/AIA. This composite was processed using the NAFE algorithm \citep{Druckmuller_2013}. The white arrows point to the twirling filaments in the prominence. The orange arrows point to the very faint envelope of the ensuing CME. This example shows the complexity of the twirling motion of the prominence as it erupted between the opposite magnetic polarities of the sunspots underlying the active region. The associated animation shows the SDO/AIA composite spanning 2011 July 06 06:03 UT to 06:42 UT with 12 s cadence (8 s total duration).}}
\end{figure}
%
Such prominence dynamics should come as no surprise given that their almost 100 times higher density and 100 times lower temperature than their immediate coronal environment results in a very sharp gradient at the prominence-corona interface, often referred to as the prominence-corona transition region, or PCTR. This sharp pressure gradient is further exacerbated by their straddling magnetic polarity inversion regions across the solar surface, where significant shear at their anchor points occurs. (See \citet{Tandberg-Hanssen_1995} for a comprehensive account of prominences.)

However, missing from the H$\alpha$ observations is the impact of prominence dynamics on their immediate coronal environment, which cannot be assessed without coronal observations.
 Taking advantage of space-based ultraviolet (UV) observations from the Solar Optical Telescope (SOT) on {\it Hinode},  \cite{Ryutova_2010} and \cite{Berger_2010,Berger_2012} were the first to independently report the formation of vortex rings following the development of a Rayleigh-Taylor (RT) instability in the vicinity of a prominence. Capitalizing on the multi-wavelength extreme ultraviolet (EUV) observations from the Atmospheric Imaging Assembly instrument \citep{Lemen2012} onboard the \textit{Solar Dynamics Observatory} (SDO), 
 \citet{Li_2012} captured the interplay between the cyclonic/tornado-like motions of a prominence eruption, akin to those reported by  \citep{Ohman_1968}, and coronal structures in their immediate vicinity (see example of Fig. \ref{SDOarrows}). 
  In addition to reports of a CME associated with the `tornado' effect, as shown in the example of Fig. \ref{SDOarrows} (see orange arrows pointing to the faint white CME envelope), \citet{Foullon_2011} and \citet{Ofman_2011} were the first to discover KH wave instabilities associated with a CME ejecta following a prominence eruption.
  
  Despite the breakthroughs achieved with these space-based observations, the collisionally excited nature of EUV emission limits its radial detection to a few 10s of Mm above the solar limb, thus limiting the exploration of prominence-induced plasma instabilities further away from the Sun.
In contrast, and despite their paucity, occurring once every 12 to 18 months, and their few minutes duration, TSE white light observations, which extend out to at least 10 $R_s$ above the limb in an uninterrupted manner, offer an outstanding opportunity to explore the radial extent of the impact of the dynamics of the prominence-corona interface.
As demonstrated with novel image processing techniques \citep{Druckmuller_2006, Druckmuller_2009} applied to high-resolution TSE white light images, \cite{Druckmuller_2014} were the first to report the presence of plasma instabilities, namely vortex or smoke-type rings, faint nested expanding loops, expanding bubbles, and twisted helical structures akin to KH wave instabilities, originating in the immediate neighborhood of prominences and extending over a field of view (FOV) almost an order of magnitude larger than that of EUV observations. Further, TSE observations captured the connectivity between prominence `threads' and overlying coronal structures, unequivocally linking plasma instabilities to the dynamic behavior of prominences \citep[see][]{Habbal_2014}.  TSE observations also captured the origination of CMEs in association with prominence eruptions, and their expansion in tandem as the prominences, forming their core and often referred to as flux ropes, remain tethered to the Sun \citep{Alzate_2017}.

A natural extension of the radial span of the TSE white light observations would have been white light imaging with the Large Angle Spectroscopic Coronagraph \citep[LASCO;][]{Brueckner_1995}, namely LASCO/C2 and C3 coronagraphs, with a continuous coverage of the corona starting at 2.5 $R_s$ and extending out to 30 $R_s$. Spanning three decades, they led to the discovery of streamer blobs \citep{1997Sheeley, 2002Sheeley&Wang, 2007Sheeley&Wang, Sheeley_2009}, which were interpreted as the relics of the helical structures of magnetic flux ropes, albeit associated with the gradual expansion of diffuse arches away from the Sun. 
\cite{Alzate_2024} extended their detection down to 0.6 $R_s$ from the solar limb by combining advanced image processing techniques \citep{Alzate_2021} with non-radial tracking of streamer boundaries \citep{Alzate_2023} applied to the SECCHI coronagraphic images on STEREO.
Unfortunately, the occulters limit the tracking of their origins down to the solar surface, and their spatial resolution further preclude them from detecting vortex rings.

Fortunately, the recent advent of the Wide-Field Imager\citep[WISPR;][]{Vourlidas2016} onboard Parker Solar Probe \citep[PSP;][]{Fox2016} which is now returning higher resolution white light coronal images, compared to LASCO and SECCHI, due to its unsurpassed proximity to the Sun, is a unique opportunity to track the evolution of the different plasma instabilities discovered with TSE observations, in particular vortex rings.
Recently \cite{Ascione_2024} reported the detection of a single elliptically shaped ``magnetic island feature" at 38 $R_s$, akin to a torus with a density depression at its center, which they assumed to be formed in-situ through magnetic reconnection in a streamer current sheet. 
\citet{Liewer_2023} also reported the presence of `blobs' with a description very similar to Ascione et al's magnetic islands. \citet{Paouris_2024} detected a trail of small-scale features along the wake of a CME between 7.5 and 9.5 $R_s$, which they associated with KH wave instabilities. However, despite their higher spatial resolution, WISPR observations still miss the direct connectivity between these reported plasma instabilities and their origin at the Sun.
In this work, we take advantage of the unique attributes of TSE and WISPR white light images, namely their spatial resolution and complementary spatial coverage, to track the evolution of vortex rings, as well as KH waves and the filamentary nature of CMEs, as they emerge from the PCTR and propagate away from the Sun.

Unless otherwise noted, we use the term vortex rings to refer to what the WISPR studies refer to as `blobs' and/or `magnetic islands'. KH wave instabilities have been detected in both. CMEs are included in this study since they are the penultimate by-products of the dynamics of the PCTR. Section 2 highlights examples from seven different TSE observations chosen at different times within two solar cycles, SC 24 and SC 25, around solar minimum and maximum. 
The salient features of the vortex rings, KH wave instabilities, and CMEs are then compared to their counterparts in the WISPR data in Section 3. 
A more quantitative assessment of the evolution of these instabilities, namely their size distribution and speeds, is given in Section 4. Section 5 concludes with the finding that the ubiquitous vortex rings, KH wave instabilities, and CMEs, originating at the inherently unstable PCTR,  propagate practically unscathed with the ambient slow solar wind.

\section{Plasma Instabilities in TSE White Light Images}
\label{tse}

The TSE white light images presented here were selected around the minimum of solar cycle 24 (SC24) (Figs. \ref{tse2019D}, \ref{tse2020B}, and \ref{tse2021C}) and around the maxima in SC24 and SC25 (Figs. \ref{tse2012}, \ref{tse2013C}, \ref{tse2023B} and \ref{tse2024}). They are representative examples from a large database spanning several decades (see \href{http://www.zam.fme.vutbr.cz/~druck/eclipse/}{http://www.zam.fme.vutbr.cz/$\sim$druck/eclipse/}). 
Every feature with a pinkish hue represents chromospheric (H$\alpha$) emission originating from prominences. Acquired by different observers, the images were processed by M. Druckm\"uller  \citep[see][]{Druckmuller_2006, Druckmuller_2009} to reveal the details of the instantaneous status of coronal structures at coronal temperatures, including dynamic events, down to the limit of the spatial resolution of the optical systems used.
In each figure, the full FoV of the TSE image is shown at the top. The different boxes within these frames, selected to highlight the different plasma instabilities captured in the images, are shown enlarged below.  Examples of vortex rings are highlighted with circles; they are the only features without annotation, given their ubiquity throughout these images. KH wave instabilities are highlighted with ellipses and labeled KH. References to CME cores and their tethers are annotated as well. Arrows are used for pointing to different features, with corresponding annotations.

\subsection{TSE Images from Solar Minimum}

\begin{figure}[!h]
 \centering
 \includegraphics [width=0.7\textwidth]{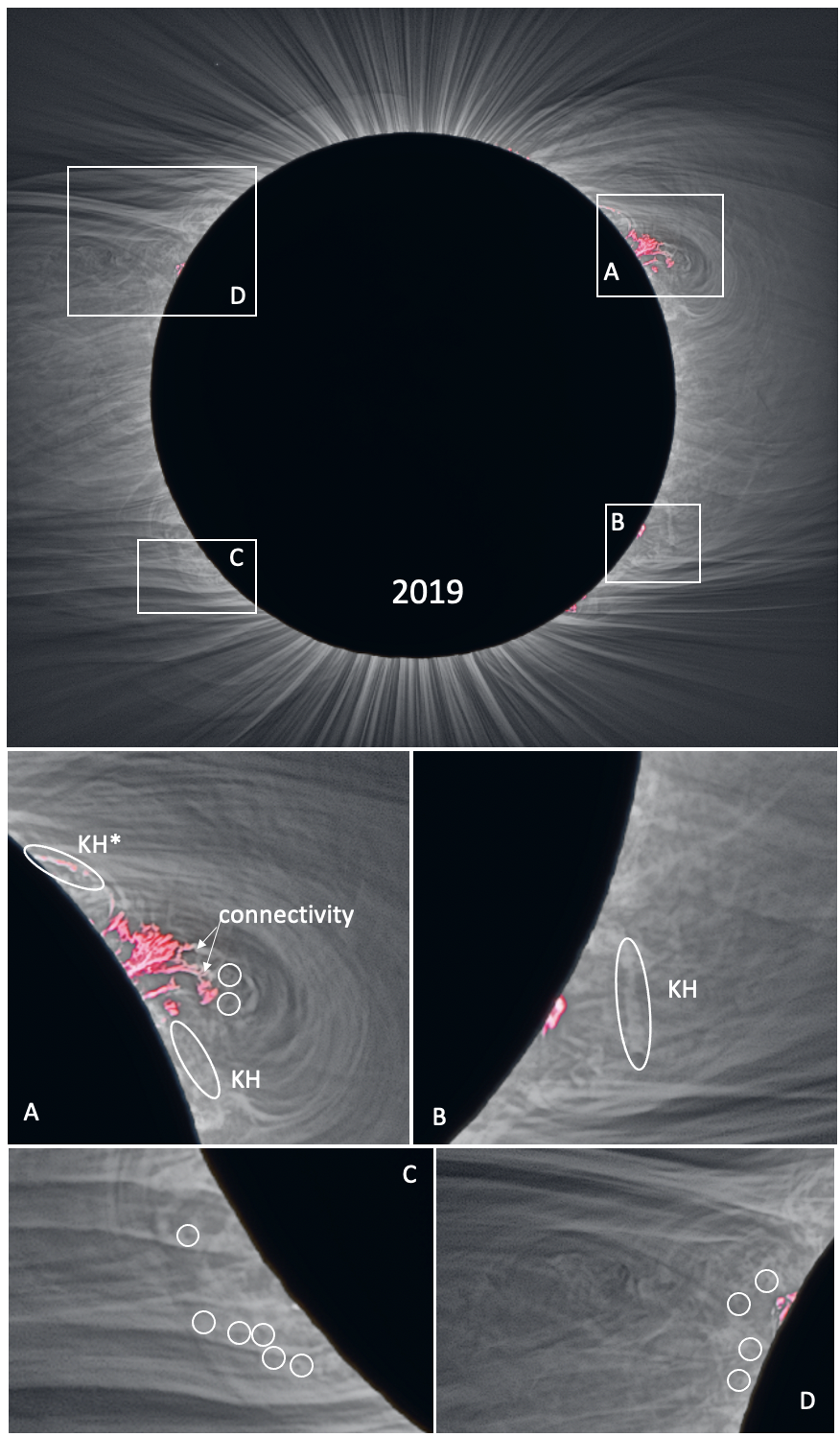}
\caption{\label{tse2019D} { \footnotesize {The 2019 July 22 TSE white light image at solar minimum, with the four enlarged panels A, B, C, and D, below. Although ubiquitous in all frames, a few vortex rings are encircled in panels A, C, and D as a guide. In panel A, a KH$^*$ at chromospheric (H$\alpha$) temperature and another at coronal temperatures (white) are encircled by ellipses. Another KH is detected in panel B. An intricate connectivity between prominence and coronal material is captured in panel A. }}}
\end{figure}

The first example shown in Fig. \ref{tse2019D} from TSE 2019 July 2,  
was taken from Tres Cruces, Chile, by Peter Aniol and Miloslav Druckm\"uller. The corona was dominated by extended polar coronal holes, while streamers hosted an abundance of vortex rings in the immediate vicinity of prominences localized at their base. Vortex rings (circles) are ubiquitous in every frame. A few are encircled in panels C and D as a reference to the eye. 
Panel A is a  particularly interesting region in the corona as it captured the connectivity between the intricate threads of a prominence (pinkish hue) and structures in its immediate vicinity at coronal temperatures (white). The two ellipses in this panel are labeled KH$^*$ at chromospheric (H$\alpha$) temperature and the other KH at coronal temperature (white) (see also another example in panel B). These are almost identical to the KH waves identified by \citet{Foullon_2011} at the flank of a fast CME captured in AIA, albeit with a higher spatial resolution than our observations (see their Fig. 1).

\begin{figure}[!h]
 \centering
 \includegraphics [width=0.8\textwidth]{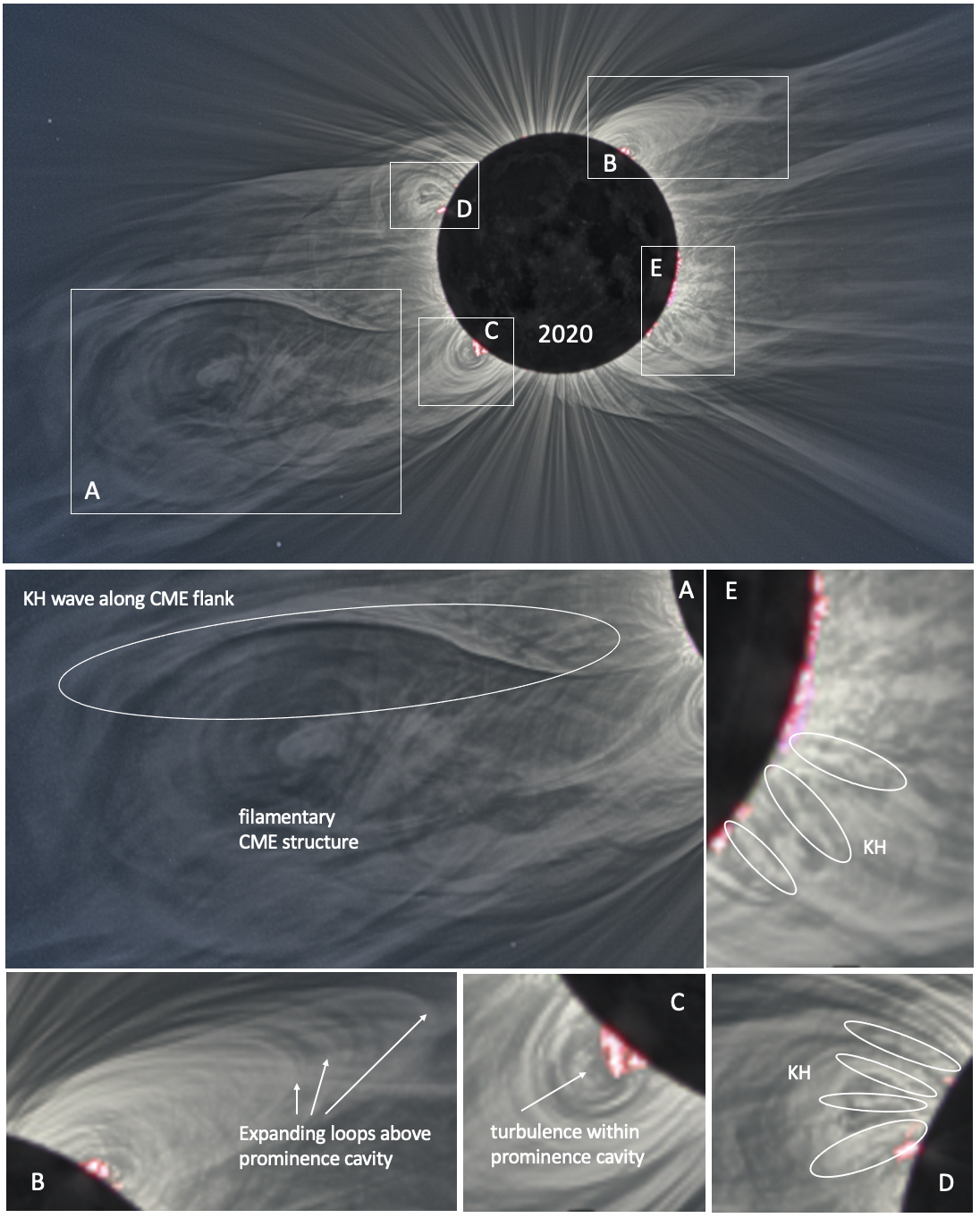}
\caption{\label{tse2020B} { \footnotesize {Top panel: The 2020 December 14 TSE white light image around solar minimum. Details of regions A, B, C, D, and E are shown in more detail below. Panel A highlights the filamentary nature of a huge CME bubble off the east limb, with a KH wave along its flank (ellipse). Panel B shows a sequence of expanding loops above the prominence cavity. Box C is a close-up of turbulence within a prominence cavity. Panels E and D captured the KH waves (ellipses).
 }}}
\end{figure}

The second example shown in Fig. \ref{tse2020B} was acquired by Andreas M\"oller from Neuquén, Argentina, during the 2020 December 12 TSE. Although much closer to solar minimum than TSE 2019, a spectacular CME spanning 3 $R_s$ above the limb, while clearly remaining tethered to the solar surface, was present during the eclipse. The complexity of the fine filamentary structure of this CME is shown in the enlarged panel A. Also captured in this panel is a KH wave, along its flank (encircled by an ellipse). The same KH waves, as first reported in Fig. \ref{tse2019D}, are present in panel E. Expanding loops above a prominence cavity appear in panel B, which are likely the precursors of a CME. The `turbulent' nature of a prominence cavity is seen in panel C, while an erupting prominence within its cavity is captured in panel D. 

\begin{figure}[!h]
 \centering
 \includegraphics [width=0.8\textwidth]{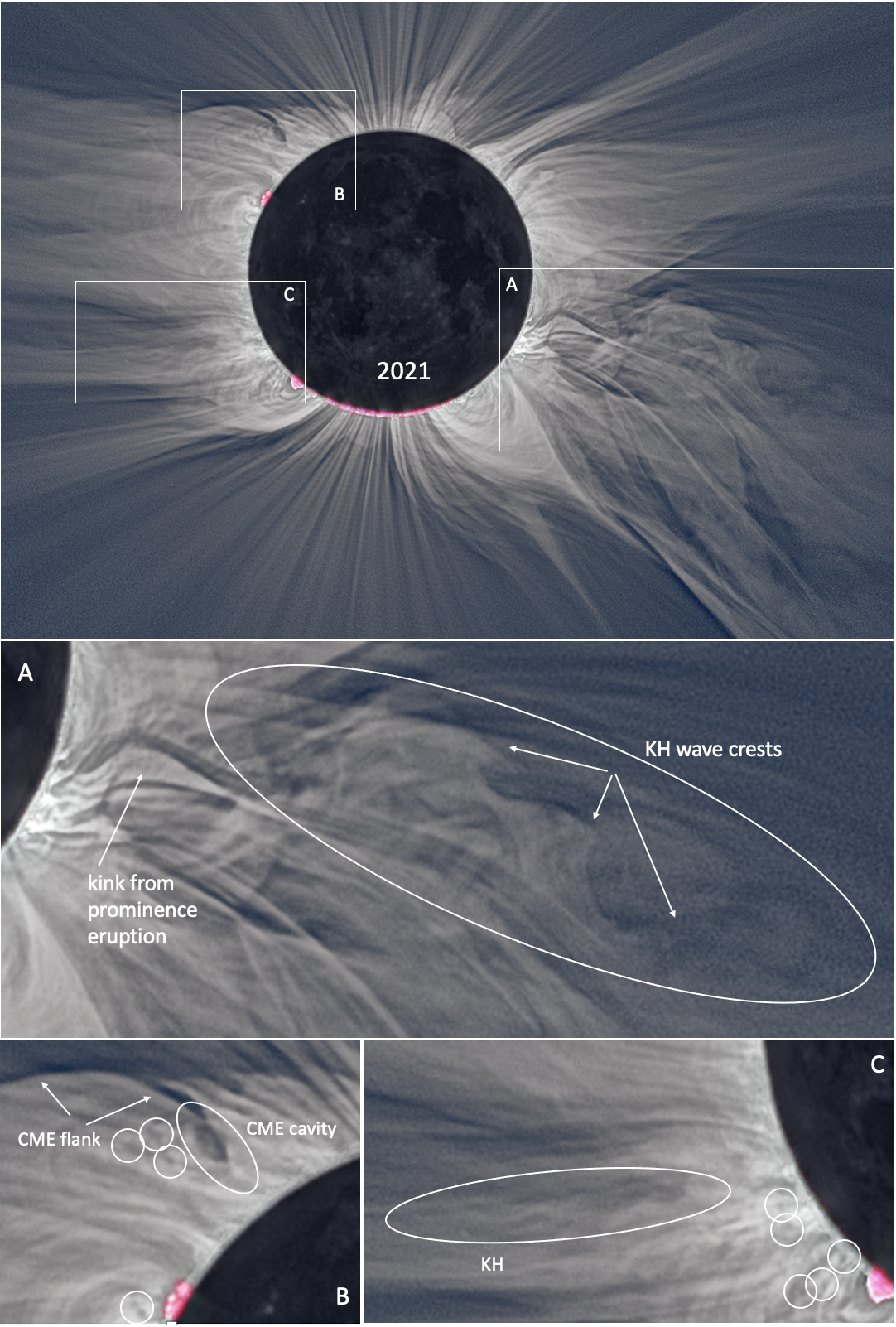}
\caption{\label{tse2021C} { \footnotesize {Top panel: The 2021 December 4 TSE white light image at 07:45 UT taken from Union Glacier in Antarctica by Andreas M\"oller around solar minimum. The details in panel A highlight a huge KH wave instability, most likely triggered by a prominence eruption followed by a CME (see arrow pointing to kink). Panel B shows vortex rings (circles) in the proximity of the CME cavity and close to the PCTR. Panel C shows another example of vortex rings in the proximity of a PCTR and KH wave instabilities.
 }}}
\end{figure}

The third example from solar minimum, shown in Fig. \ref{tse2021C}, was acquired by Andreas M\"oller from Union Glacier in Antarctica during TSE 2021, December 4. It shows a spectacular KH wave instability off the west limb. The LASCO\href{https://cdaw.gsfc.nasa.gov/CME_list/}{\footnote{https://cdaw.gsfc.nasa.gov/CME$\_$list/index.html}} 
catalog indicates that a CME had emerged from the inner edge of the LASCO/C2 occulter at 05:24 UT, at position angle (PA) = 270$^o$. The CME, labeled a `partial halo' in the LASCO catalog, with a speed of 523 km/s, was most likely triggered by the prominence eruption indicated by the arrow pointing to a kink in panel A. Like the other previous examples, there is a plethora of vortex rings within the streamers off the east limb (see panel B). Another much smaller KH wave instability off the east limb around PA = 100$^o$, highlighted in panel C, was most likely associated with the same CME.

\subsection{TSE Images from Solar Maximum}

\begin{figure}[!h]
 \centering
 \includegraphics [width=0.9\textwidth]{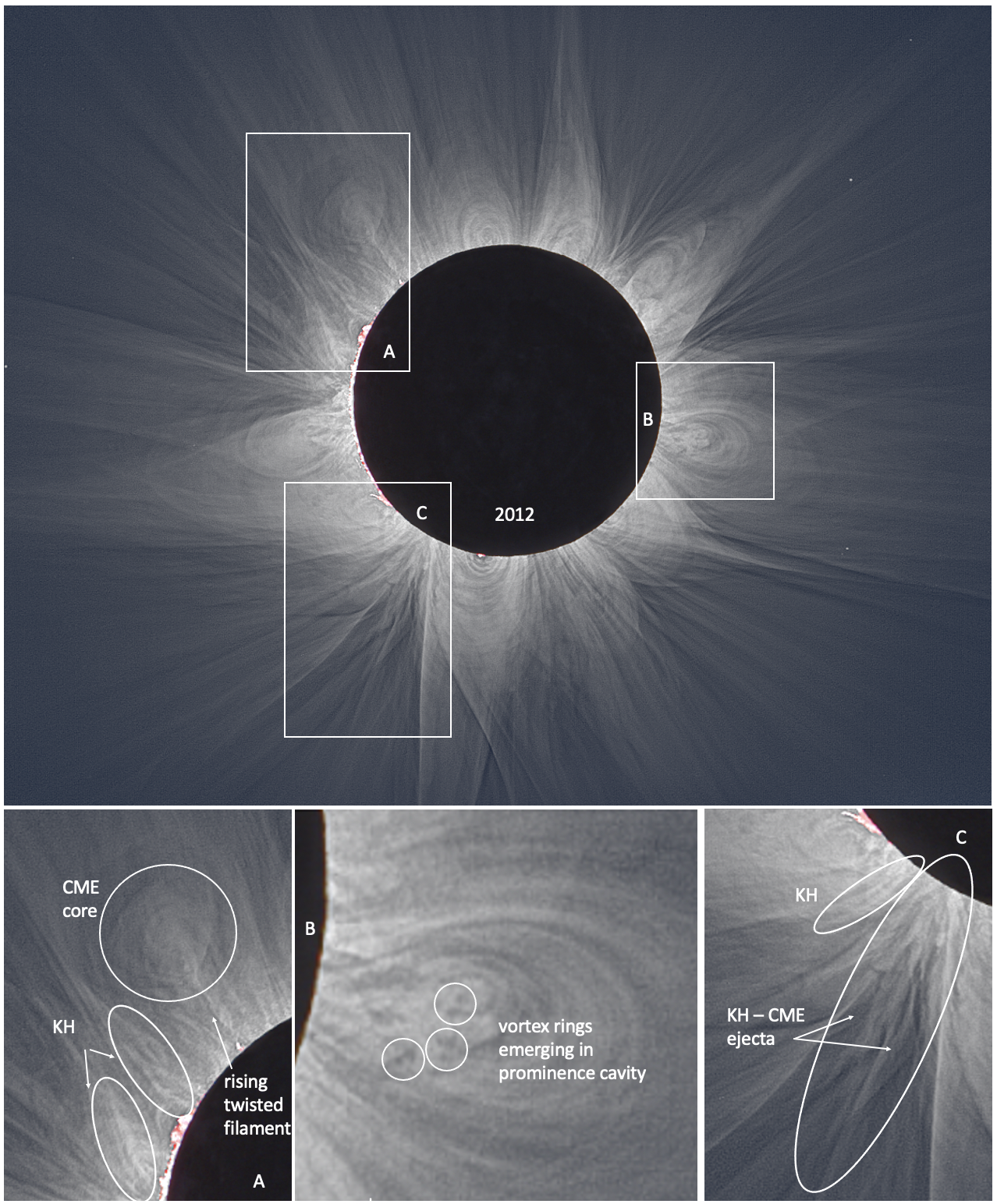}
\caption{\label{tse2012} { \footnotesize {The 2012 November 13 TSE white light image at solar maximum, acquired by Constantinos Emmanoulidis. Panel A shows a rising twisted filament with adjoining KH waves (enclosed within ellipses). In panel B, vortices are seen emerging in a prominence cavity. 
Panel C is an example of KH instabilities similar to those found in Fig. \ref{tse2021C}A, albeit with smaller amplitudes. 
 }}}
\end{figure}

\begin{figure}[!h]
 \centering
 \includegraphics [width=1.0\textwidth]{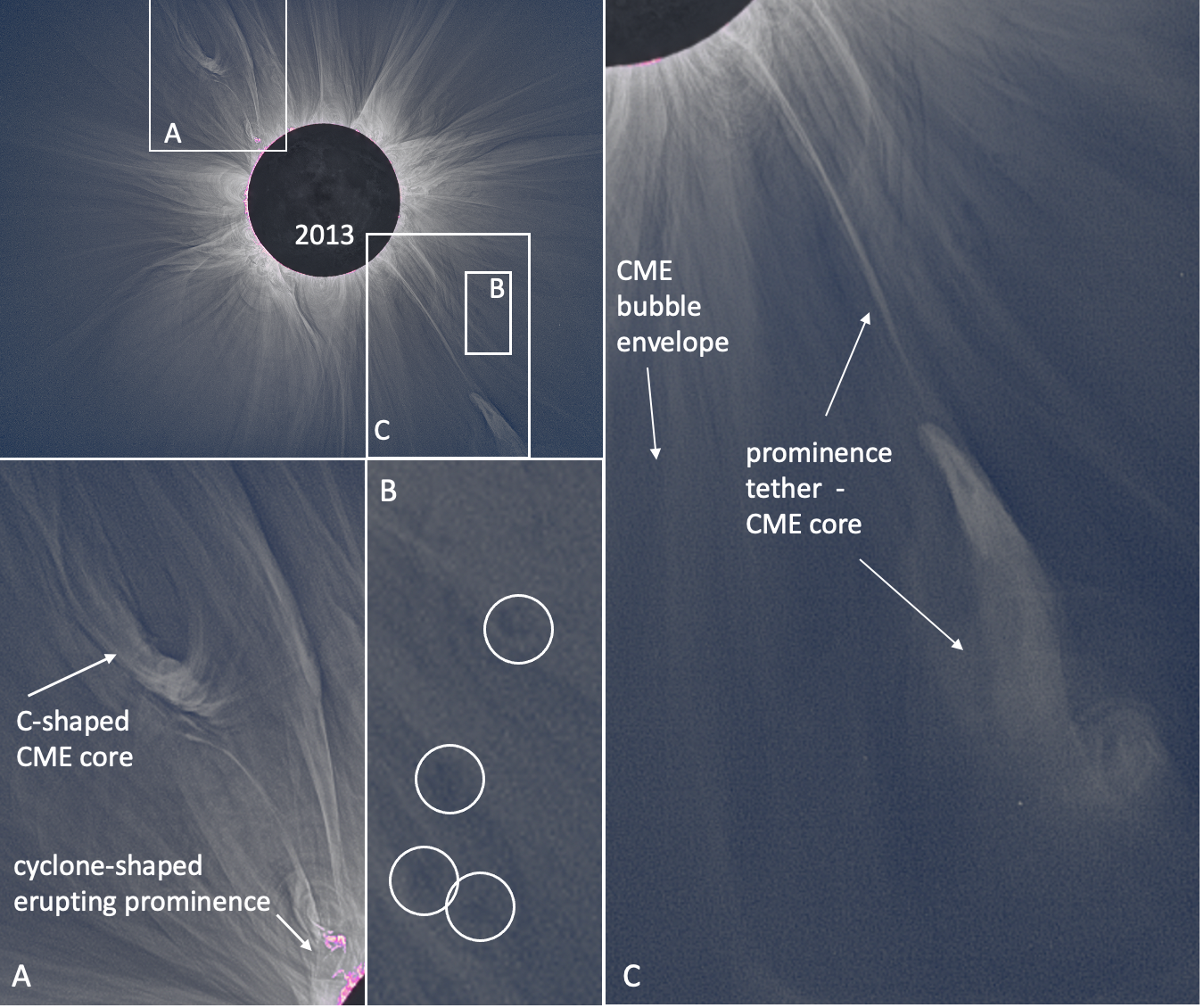}
\caption{\label{tse2013C} { \footnotesize {The 2013 November 3 TSE white light image at solar maximum, acquired by Constantinos Emmanoulidis. Panel A highlights a C-shaped CME core produced by a filament eruption. Panel B highlights faint, vortex rings at 2 $R_s$, which appear to be embedded in coronal ray-like structures. Panel C shows the extended field of view of the overview image, highlighting the CME/prominence core in the shape of a question mark, still tethered to the Sun, with a faint CME bubble envelope. 
 }}}
\end{figure}

\begin{figure}[!h]
 \centering
 \includegraphics [width=1.0\textwidth]{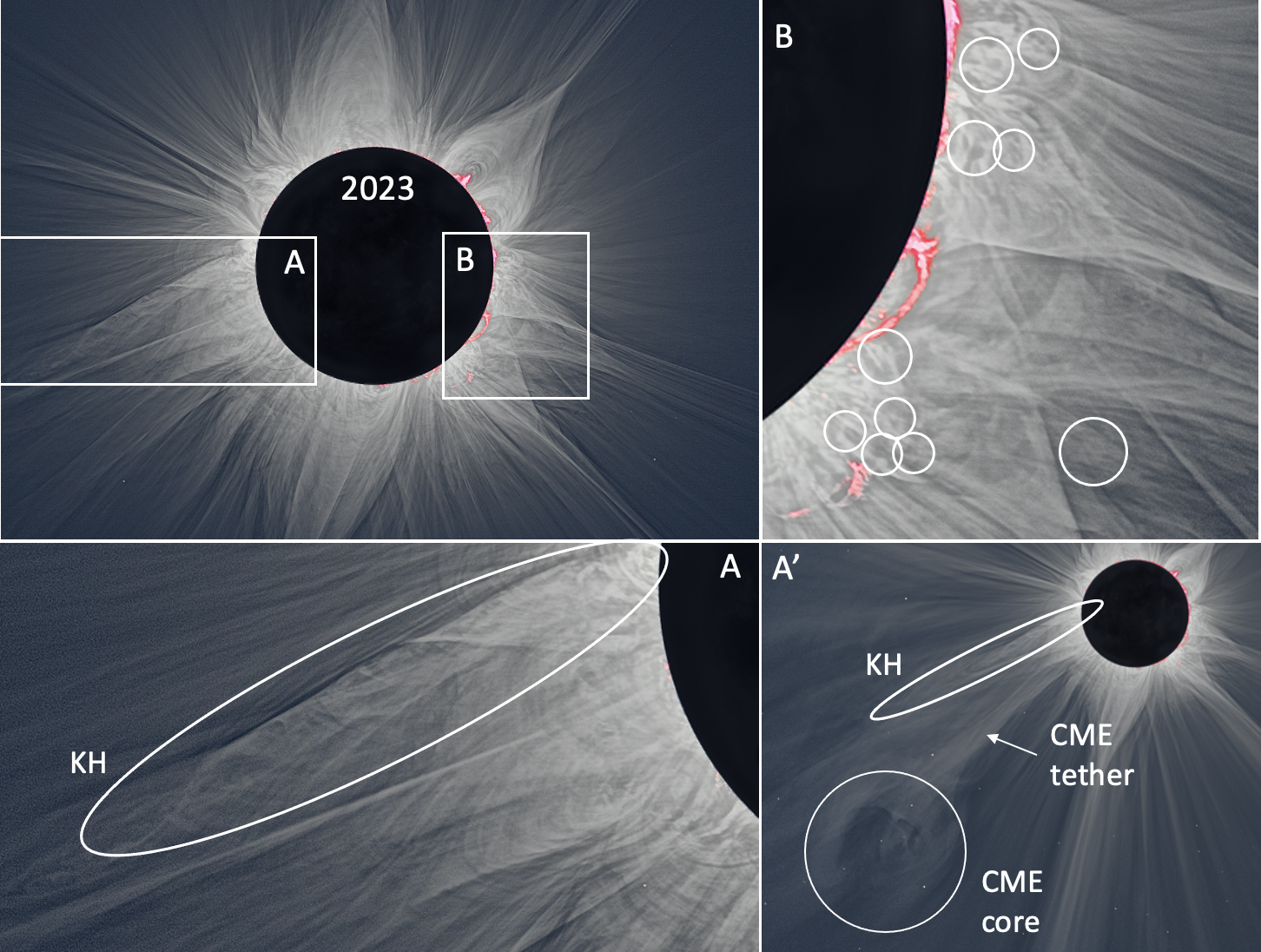} 
\caption{\label{tse2023B} { \footnotesize {The 2023 April 20 TSE white light image at solar maximum. Panel A is an enlarged section of A', highlighting the details of the KH wave instability that developed in the wake of the faint CME shown in the wider field of view (A'). Panel B highlights a number of vortex rings. 
 }}}
\end{figure}

\begin{figure}[!h]
 \centering
 \includegraphics [width=0.9\textwidth]{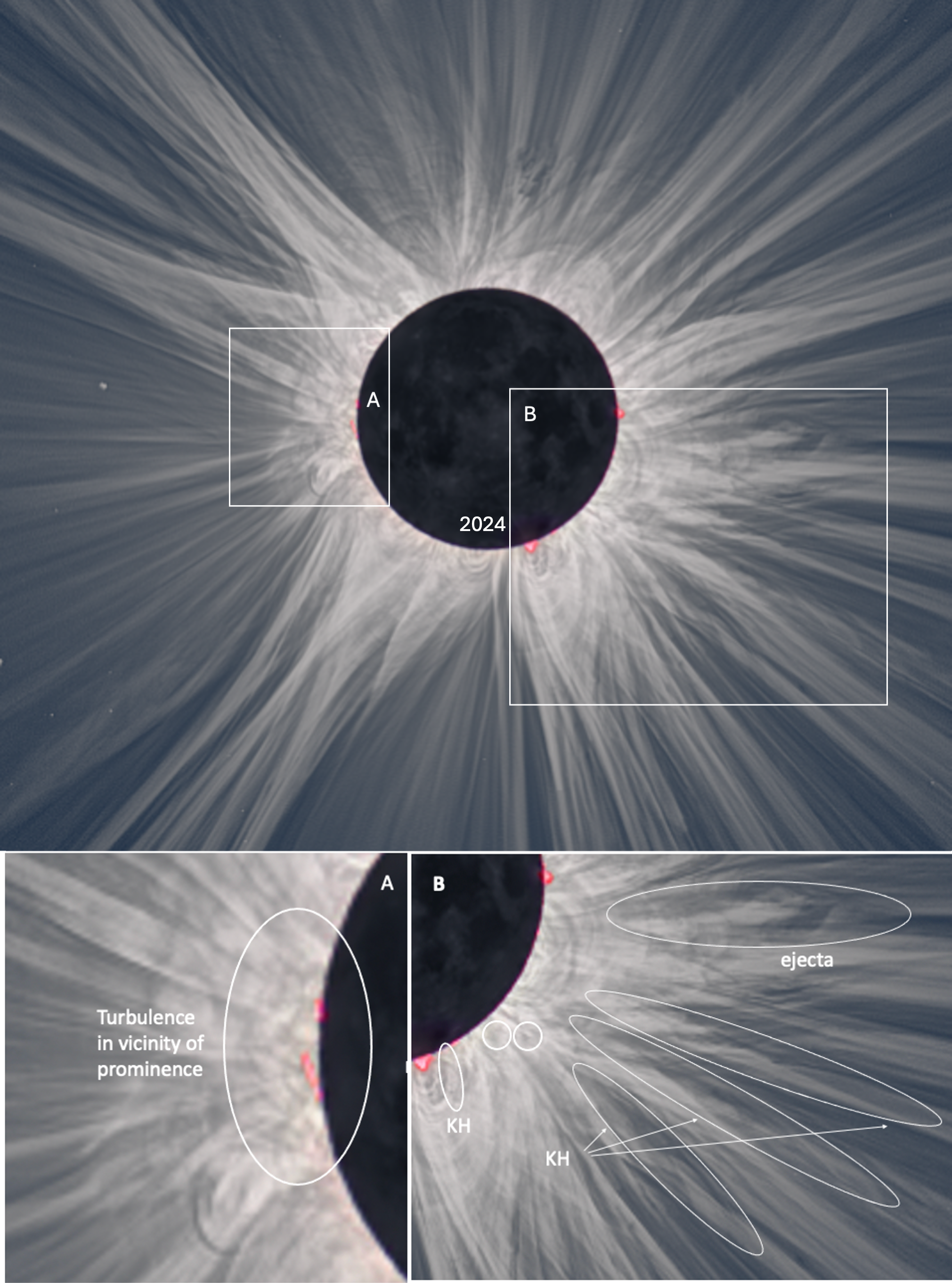}
\caption{\label{tse2024} {The 2024 April 8 TSE white light image close to solar maximum in SC25. A clearly turbulent corona appears above low lying prominences at PA = 90$^o$, shown encircled by an ellipse in panel A. Several KH wave instabilities and clear ejecta were captured off the west limb. A low lying KH wave appears along the flank of low-lying rising loops above a very bright prominence similar to the examples in Figs. \ref{tse2019D} to \ref{tse2012}. Examples of a few vortex rings are given in panel B.
}}
\end{figure}

The examples around solar maximum from SC24 (2012 and 2013) and SC25 (2023 and 2024), are shown in Figs. \ref{tse2012}, \ref{tse2013C}, \ref{tse2023B} and \ref{tse2024}. The distribution of large-scale structures, namely streamers, differs significantly from the solar minimum examples, with prominences and streamers appearing at all latitudes. This is particularly striking in the examples of 2012 and 2013. 

The  2012 November 13 TSE image shown in Fig.~\ref{tse2012} was acquired by D. Finlay and C. Emmanoulidis from Northern Australia. While this image has the lowest spatial resolution among all the TSE examples, it captured several examples of rising bubbles, all around the northern hemisphere, most likely precursors to CMEs, associated with rising filaments. Details of the ones entraining the overlying closed structures with them are shown in panel A. A preponderance of emerging vortex rings within a prominence cavity are shown in panel B. Similar to the examples of Figs. \ref{tse2019D} and \ref{tse2020B}, there are several KH waves along the flanks of the rising structure in panel A. KH-CME associated waves, similar to wakes, are captured in panel C following the CME ejecta.

The second example from SC24 maximum  shown in Fig. \ref{tse2013C} was acquired by  C. Emmanouilidis on 2013 November 3 from Gabon. (It was studied in detail by \cite{Alzate_2017}, albeit with emphasis on the role of CMEs in shaping coronal structures in their path.) The unsurpassed spatial coverage of this image, and its southern extension shown in panel C, spanned several solar radii starting from the limb. These images captured two examples of tethered CMEs, with their bright prominence core and much fainter envelopes, as shown in panels A and C. Panel A further shows the details of a C-shaped CME core. Panel B reveals the presence of faint vortex rings captured at 2 $R_s$ further north of the tethered CME in panel C. 

The final two examples from SC25 present two strikingly different coronae at solar maximum. The 2023 April 22 TSE image Fig. \ref{tse2023B} was taken by Pavel Starha from the Ningaloo peninsula in Northwest Australia. The extension of the FoV is shown in panel A', where a very faint CME appears around PA = 130$^o$, about 5 $R_s$ above the limb. It was accompanied by a KH wave instability trailing it to the north, as shown in the enlarged panel A, extending out to at least 2 $R_s$ above the limb. Shown in panel B is a plethora of vortex rings between PAs 180$^o$ and 270$^o$, clearly positioned above a series of raised and `suspended' prominences above the limb.

The 2024 April 8 TSE image taken by Petr Starha and Jana Hoderóva from Sims, Arkansas, is shown in Fig. \ref{tse2024}. It was a very bright corona dominated by ray-like structures distributed almost spherically symmetrically around the limb. Panel A highlights the turbulent appearance of the immediate neighborhood of a few prominences at the east limb at PA = 90$^o$. Several KH wave instabilities and clear ejecta were captured off the west limb following the wake of a CME. A very low lying KH wave appears along the flank of low-lying rising loops above a very bright prominence (which was clearly visible to the naked eye), similar to the examples in Figs. \ref{tse2019D} to \ref{tse2012}. Examples of a few vortex rings are given in panel B. No CMEs were captured in the FOV during totality, and curiously, none were reported in the LASCO/C2 catalog.

\section{Comparison of Plasma Instabilities in TSE and WISPR white light images}

The ensemble of TSE images shown in Figs. \ref{tse2019D} through \ref{tse2024} captured a plethora of vortex rings independently of the phase of any of the two solar cycles considered. While the largest concentrations appeared in the immediate vicinity of prominences, namely the PCTR, their spatial distribution throughout the corona is the most direct proof we have from these still images that vortex rings are far from stationary, essentially, that the TSE images are capturing their instantaneous location at any given time.  
While the dynamics of KH wave instabilities and CMEs have been amply documented from space-based observations, vortex rings are novel findings. Fortunately, the recent observations from WISPR/PSP lend support to their evolutionary status. While non-contemporaneous, the temporal evolution of plasma instabilities ranging from vortex rings to KH wave instabilities and CMEs in the WISPR data, to be presented next, is essential for exploring how these different manifestations of plasma instabilities originate at the PCTR and evolve through the extended corona into interplanetary space.

\subsection{WISPR/PSP data overview}

WISPR is the only remote sensing instrument onboard PSP designed to image the solar corona and inner heliosphere over a wide FOV in white light.  It comprises two wide-angle telescopes, the inner telescope (WISPR-I) observes solar elongations ranging from $\sim 13.5^{\circ}$ to $53.5^{\circ}$, while the outer telescope (WISPR-O) extends the coverage from ${\sim} 50^{\circ}$ to $108^{\circ}$. As PSP orbits the Sun with a rapidly changing heliocentric distance, the FOV of each detector projects onto varying distances throughout the orbit. The motion of the spacecraft as it approaches or recedes from perihelion results in a dynamic spatial coverage of the inner heliosphere across different solar encounters. 
The spectral bandpasses of WISPR-I and WISPR-O are $490–790$ nm and $475–795$ nm, respectively, enabling broadband white-light imaging. In the inner heliosphere, WISPR captures white-light emission from the photospheric light scattered by free electrons \citep[i.e., the K-corona;][]{Billings1966} known as Thomson scattering, as well as by interplanetary dust particles orbiting the Sun \citep[i.e., the F-corona;][]{Grotrian1934, Kimura_1998}. 

A variety of dynamic coronal structures, i.e., characteristically different from the background solar wind, have already been observed by WISPR, as will be shown in the examples of Figs. \ref{tse-wispr_vortex_fin}, \ref{tse-wispr_KH_fin}, and \ref{tse-wispr_cme_fin}. The prominent features are CMEs with their complex internal structures shown in Fig. \ref{tse-wispr_cme_fin}. Recent studies by \cite{Wood_2023, Shaik_2024} and  \cite{Cappello_2024} revealed intricate internal structures that deviate from the traditionally {\it assumed} three-part CME structure of leading-edge, cavity, and core \citep{IllingH_1985, Vourlidas_2013}, based on {\it a combination of spatially interrupted} observations from ground-based and space-based coronagraphic instruments. As demonstrated in the TSE examples presented earlier, CMEs are far more complex. Although less visually distinct compared to CMEs, vortex ring-like features, as shown in Fig. \ref{tse-wispr_vortex_fin}, are frequently observed in WISPR images \citep{Ascione_2024, howard_2019, Liewer_2024, Liewer_2023, Hess_2020}. Similarly, signatures of KH waves, akin to the TSE examples presented earlier,  have also been reported in WISPR images (see \citet{Paouris_2024}) as shown in Fig. \ref{tse-wispr_KH_fin}.

To enhance the details  of vortex rings, KH wave and CMEs in the WISPR observations we used Level 2 unprocessed calibrated data in units of Mean Solar Brightness and applied the Bandpass Frame Filtering (BFF) method \citep{Alzate_2021}. This method utilizes a wide and a narrow Gaussian filter in time to isolate brightness fluctuations between $\approx45$ min and $\approx10$ hours.  A histogram equalization is then applied to enhance contrast in the processed images. For some observations with very high brightness contrast, such as CMEs, a background brightness model (WISPR Level 2b data) is first applied before applying the BFF method.

\subsection{Vortex rings,  KH wave instabilities and CMEs in TSE and WISPR white light images}

\begin{figure}[!h]
 \centering
\includegraphics[width=1.0\textwidth]{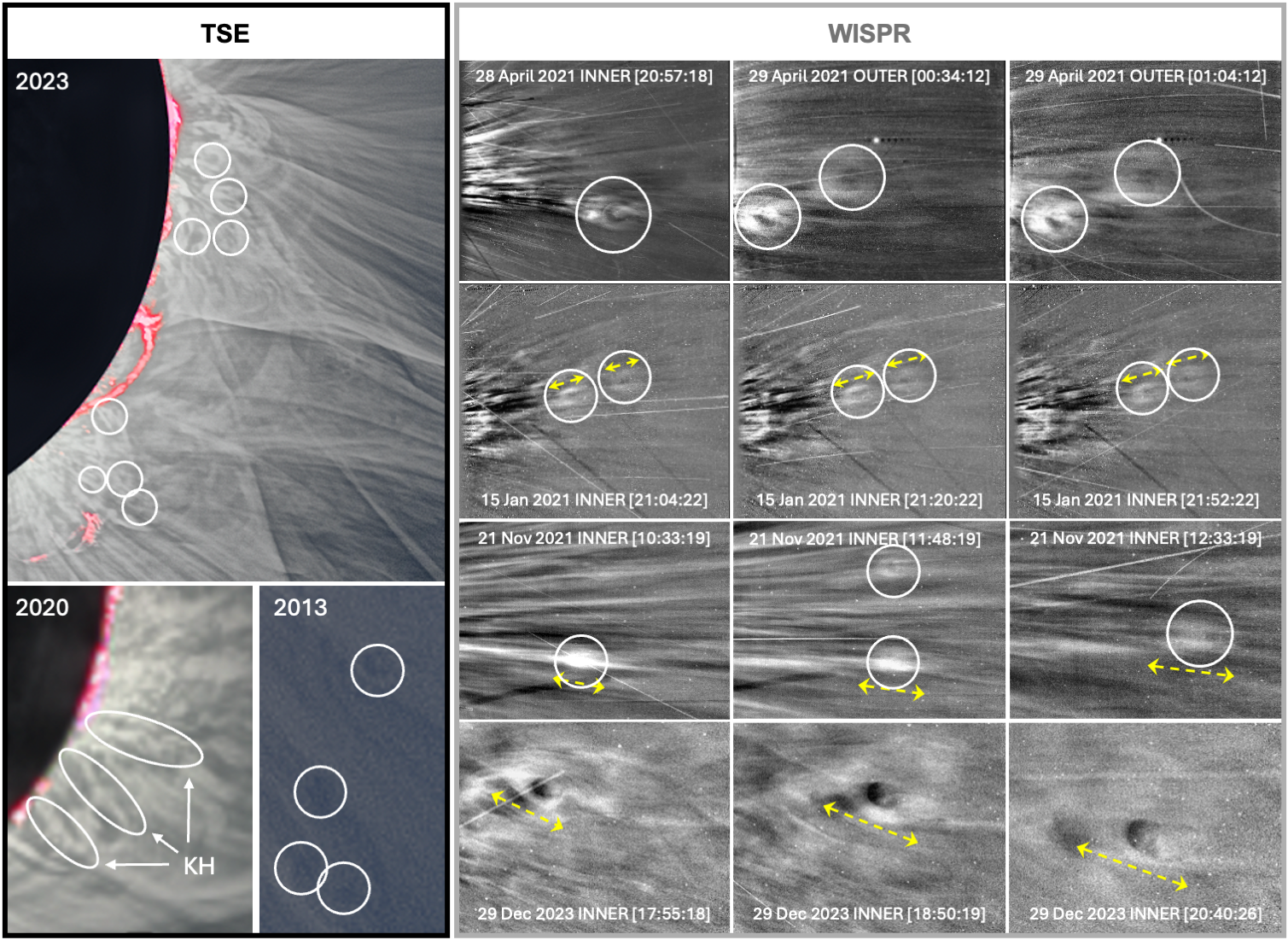}
\caption{\label{tse-wispr_vortex_fin}{Comparison of vortices in TSE (in black box) and BFF-processed WISPR (in gray box) images. The WISPR images in the right panels show various cases of vortex ring observations (on the dates marked in each panel). These include cases both with and without dark cavities at the center of the vortex rings. The associated animation (11 s total duration) shows PSP/WISPR images from 2021 April 28 16:12 UT to 29 06:04 UT (inner and outer imagers at 15 and 30 min cadence), 2021 January 15 12:16 UT to 21:20 UT (inner imager at 16 min cadence), 2021 November 21 09:03 UT to 15:18 UT (inner imager at 15 min cadence), and 2023 December 29 15:05 UT to 22:15 UT (inner imager at 5 min cadence). The white boxes in the animations indicate the zoomed-in area shown in the rightmost column.}}
\end{figure}

\begin{figure}[!h]
 \centering
 \includegraphics[width=0.7\textwidth]{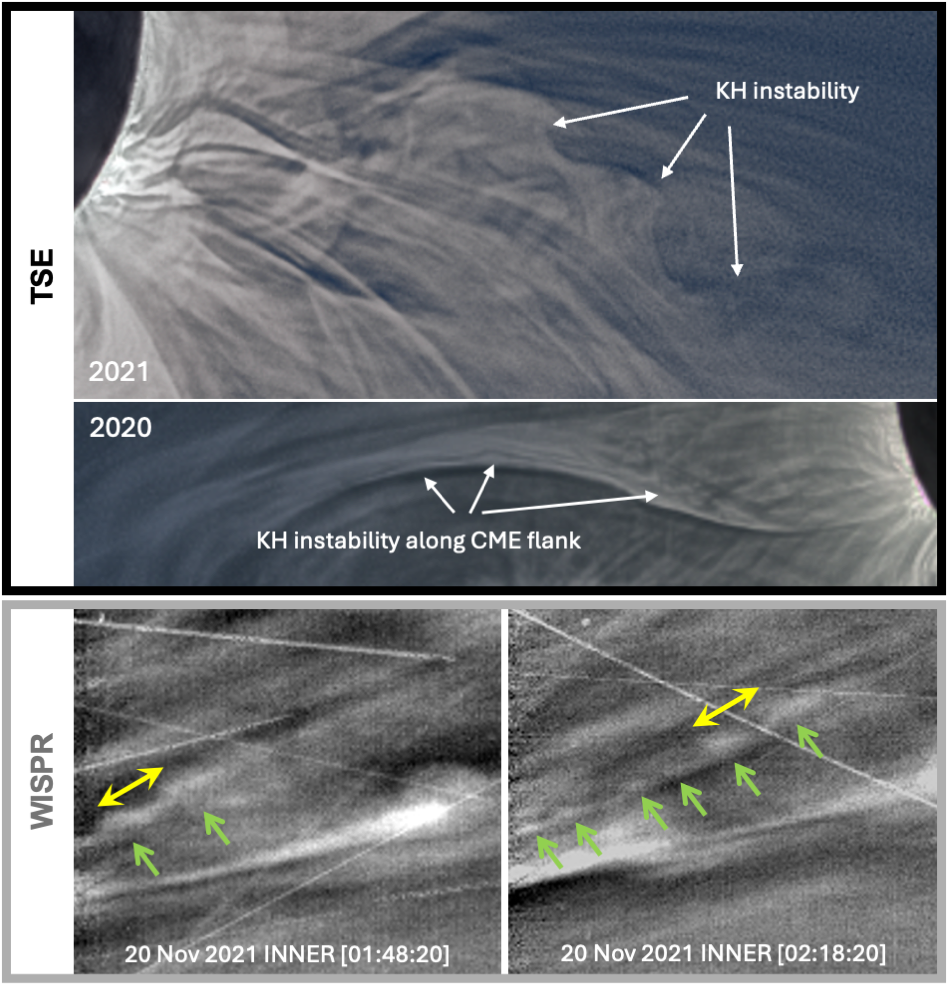}
\caption{\label{tse-wispr_KH_fin} {\footnotesize{Comparison of KH instabilities between TSE (white arrows referring to their crests) and BFF-processed WISPR images (with green arrows pointing to their crests). The yellow double-arrows in WISPR show the size scale measured between crests. The associated animation (2 s total duration) shows PSP/WISPR images from 2021 November 19 23:03 UT to 20 05:48 UT (inner imager at 15 min cadence). The white box in the animation indicates the zoomed-in area shown in the rightmost column.}}}
\end{figure}

\begin{figure}[!h]
 \centering
 \includegraphics [width=0.85\textwidth]{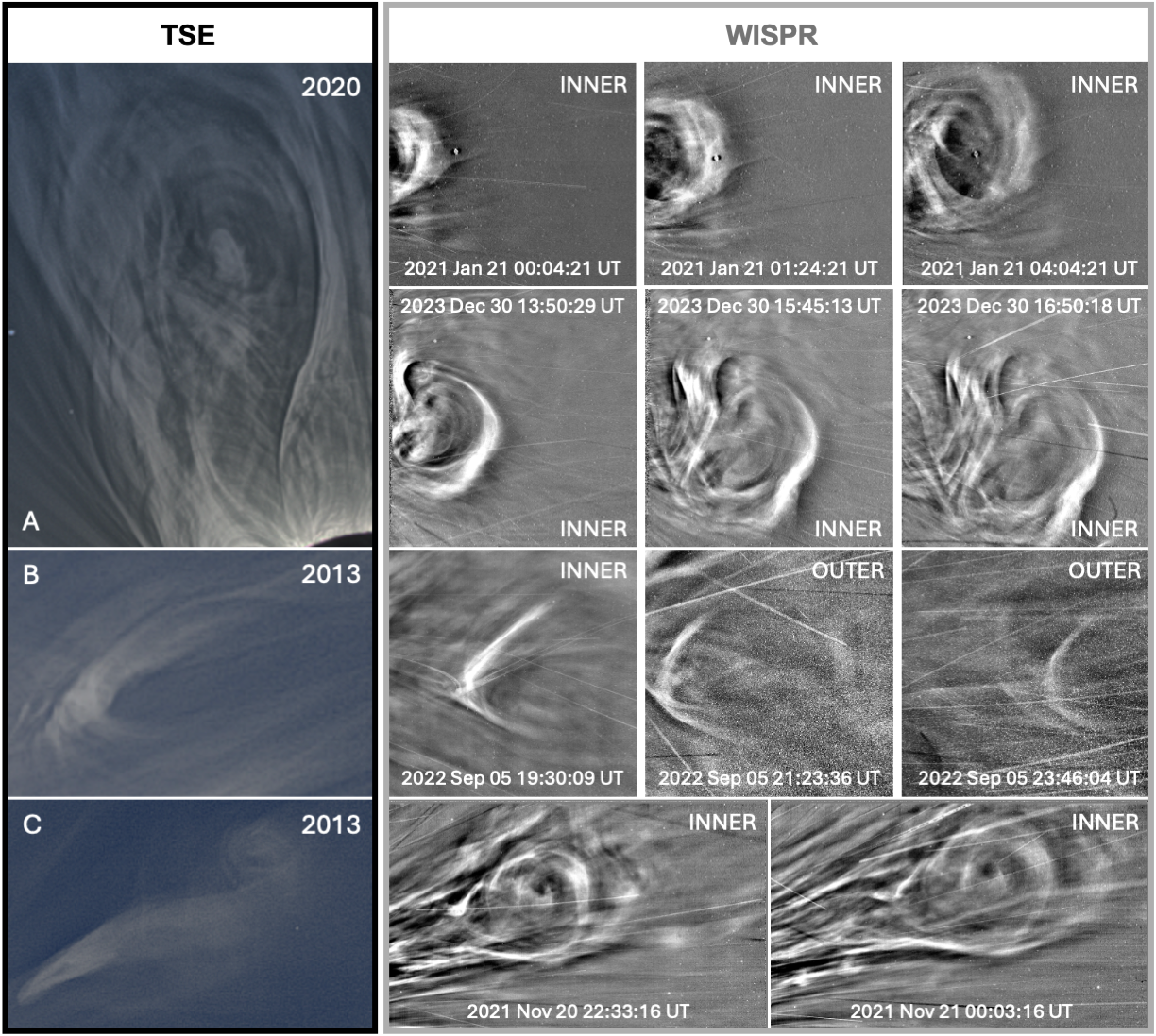}
\caption{\label{tse-wispr_cme_fin}{Comparisons of internal structures of three different types of CME cores observed in TSE and BFF-processed WISPR images, underscoring the intricate details of their filamentary substructures starting in the corona and propagating into the heliosphere. Box A depicts the more common appearance of CME cores. Box B is an example of a C-shaped CME core. Box C depicts CME cores akin to a question mark. The associated animation (17 s total duration) shows PSP/WISPR images from 2021 January 20 21:40 UT to 21 13:40 UT (inner imager at 16 cadence), 2023 December 30 10:00 UT to 19:55 UT (inner imager at 5 min cadence), 2022 September 05 17:15 UT to 06 00:38 UT (inner and outer imagers at 15 min cadence), and 2021 November 20 18:03 UT to 21 05:18 UT (inner imager at 15 min cadence). The white boxes in the animations indicate the zoomed-in area shown in the rightmost column.}}
\end{figure}

For comparison with TSE images, we selected a few examples of WISPR white light observations from several PSP solar encounters (namely, 1, 7, 8, 10, 13, and 18), highlighting features within a FoV of 5 - 15 $R_s$. They include the three different types of plasma instabilities reported in the TSE images, namely vortex rings, KH wave instabilities, and CMEs. In Figures~\ref{tse-wispr_vortex_fin},~\ref{tse-wispr_KH_fin}, and ~\ref{tse-wispr_cme_fin}, close-up examples from TSE images selected from Figs. \ref{tse2019D} to \ref{tse2024} are displayed within a black frame, while comparable structures observed in WISPR are displayed within a gray frame. The corresponding animations are available in the online version of this article. 

The first comparison shown in Fig. \ref{tse-wispr_vortex_fin} provides examples of the most widespread plasma instability, namely vortex rings, selected from TSEs 2013, 2020, and 2023. They depict a variety of individual vortex rings (2023), KH wave instabilities in the low corona (2020), as well as vortex rings captured at $\sim$2 $R_s$ (2013). The corresponding panels from WISPR, shown to the right, offer a time sequence of a few hours. Visually, WISPR vortex rings do not seem to undergo any change in shape as they expand, as shown in the top two rows. A bright vortex is captured in the third row. In the bottom row, there are two seemingly closely connected vortices (most likely a line of sight effect). The yellow dashed arrows indicate their measured elongation (discussed in the next section). A comparison of vortices in the TSE images and their WISPR counterparts reveals that they have identical shapes. The distinguishing feature is the enhanced contrast between their center region (in the bottom row of the WISPR panels), referred to as cavity, and the bounding bright boundary or `ring' similar to the ones discussed in \citet{Ascione_2024}, and \citet{howard_2019}. 

The next examples shown in Fig. \ref{tse-wispr_KH_fin} are a comparison of different manifestations of KH wave instabilities between the two data sets. Here, the largest KH wave instability from TSE 2021, shown at the top, appeared after the passage of a CME through the corona. In TSE 2020, the KH instability appeared along the CME flank. In the WISPR examples from 2021 November 20, taken about 30 minutes apart, the KH `crests', also along a CME flank, appear fainter than their TSE counterparts.

The final comparison pertaining to CMEs is shown in Fig. \ref{tse-wispr_cme_fin}. The complex filamentary nature of CMEs revealed in the inner corona from TSEs 2020 and 2013 is clearly present in the `corresponding' WISPR examples shown to their right. The intertwined structures that form the CME bubble in the TSE 2020 example (panel A) remain tethered to the Sun. In TSE 2013, the  internal structures of the two CMEs present themselves differently, most likely due to their orientation along the line of sight of the observer. (B) appears as a C-shape dip, while (C) appears as a question mark. Their `counterparts' in WISPR are surprisingly similar, with the important exception that their connectivity to the Sun cannot be established due to the limited WISPR FoV. The top two rows of WISPR are examples associated with panel A from TSE 2020, of the more classical CME type, but with complex internal structures. A C-shaped CME cusp is shown in row B of WISPR, almost identical to the one from TSE 2013. A CME with a question mark core is shown in row C with its very close counterpart from TSE 2013.

\section{Evolution of Plasma Instabilities with Radial Distance}

We show in this section how the inference of the size distribution of the same plasma instabilities detected in both TSE and WISPR data, presented above, yields a novel opportunity for exploring their evolution as they propagate away from the Sun. We rely on the assumption that the spatial distribution of plasma instabilities as a function of radial distance from the different eclipse years can be considered as their instantaneous distribution in time of their expansion throughout the corona. The availability of their temporal evolution in the WISPR data further enables us to infer their speeds, with particular emphasis on vortex rings.

\subsection{Size distribution versus distance}

\begin{figure}[!h]
\centering
\includegraphics [width=1.0\textwidth]{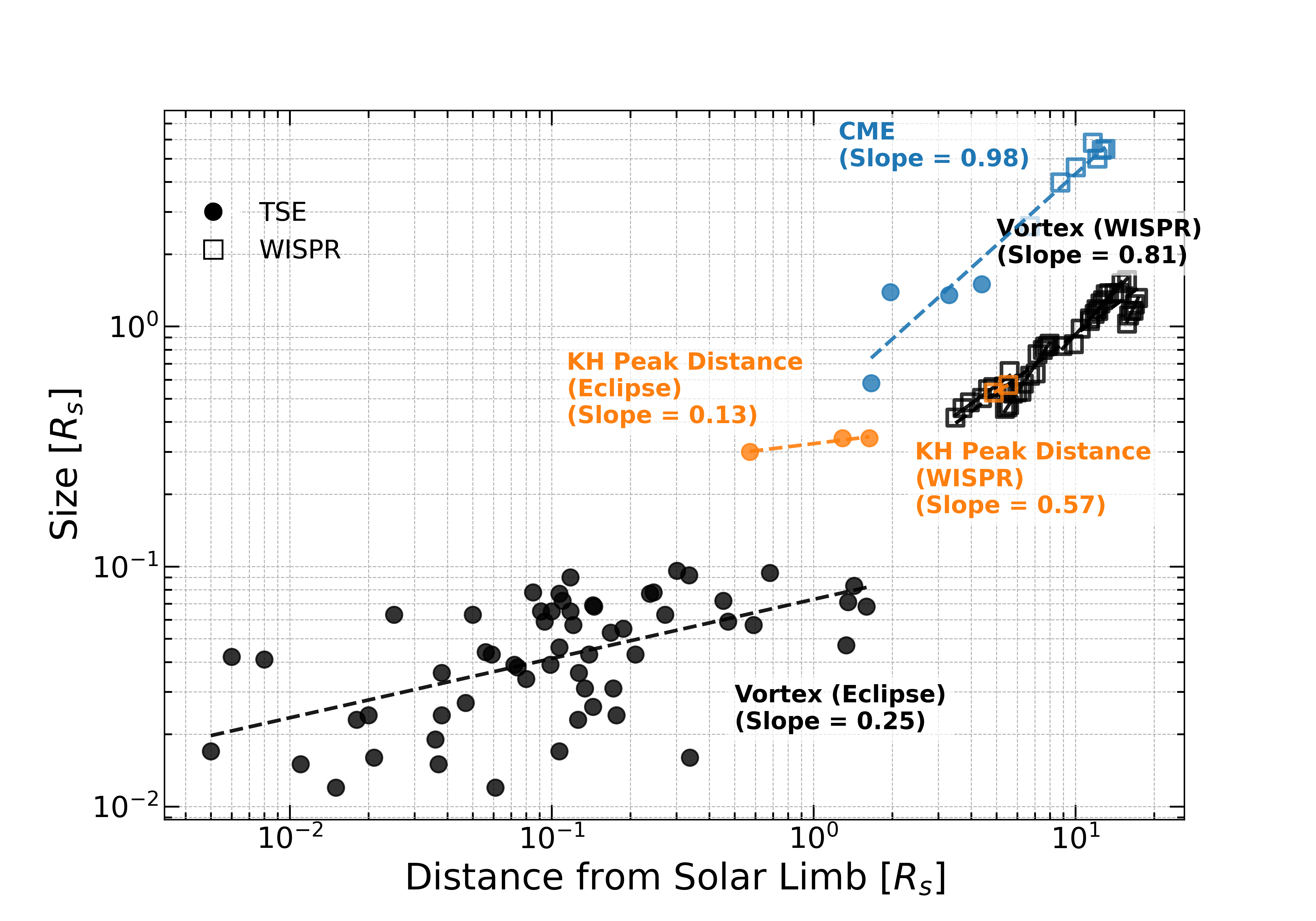}
\caption{\label{PlotR}
{\footnotesize {Size (in $R_s$) distribution versus helio-projective distance (in $R_s$) calculated for all TSE images (filled circles) and LW-processed WISPR observations (open squares). Black is for vortex rings, orange for KH waves, and blue for CMEs. The dashed lines are the best fits to the data, with their respective slopes given in parentheses. The larger TSE vortex ring sample produces a larger scatter compared to the WISPR data. }}}
\end{figure}

Similar approaches were used to measure the size distribution of the different plasma instabilities in the TSE and WISPR data. For the TSE observations, each identified vortex ring was bound by an ellipse, whose semi-major axis was measured in pixels, then converted to $R_s$, and was taken to be their size. The radial distance of each vortex ring was measured from the solar surface to the closest edge of the ellipse. The same ellipse-fitting method was applied to the four  CMEs reported in this study, using  their envelope for size, and the distance from the solar surface to the CME's core as their radial distance.
For the three KH waves, the distance between the wave crests was taken as a measure of their size, and the distance of the crest closest to the solar limb  was their radial distance.  This approach could not be applied to the miniature KH wave instabilities observed very close to the solar surface due to the spatial resolution in the TSE images, which was not sufficient to resolve the spatial separation of their crests. The resulting distribution of size versus distance is given in the log-log plot of Fig. \ref{PlotR} as filled circles, with black for vortex rings, orange for KH waves, and blue for CMEs. The best-fit slopes for each category are given in parentheses, with 0.25 for vortex rings, 0.13 for KH waves, and 0.98 for CMEs. While vortex rings and KH waves seem to have comparable expansion factors with distance, CMEs expand by almost an order of magnitude faster.

For the WISPR data, the images were first processed with the algorithm described in Appendix A of \citet{Howard_2022} (LW-processed images). The procedure effectively removes the F-corona signal and suppresses the contribution of quasi-static K-corona features such as coronal streamers. This enhancement also increases the relative contrast of the fine-scale structures in the image. To estimate the size and radial distance of each feature observed in the WISPR images, we assumed that the feature lies in the plane of sky. This assumption was adopted because of the rapid motion of the spacecraft during the observations. To minimize projection effects and uncertainties associated with Thompson scattering geometry \citep{Vourlidas_2006, Nistico_2020, howard_2019}, we preferentially selected consecutive time frames where the features were observed close to the inner boundary of the WISPR-I FoV. The sizes of the features were estimated by measuring the pixel distance between their nearest and farthest visible edges to the solar limb, as indicated by the yellow dashed lines of WISPR image panels in Fig. \ref{tse-wispr_vortex_fin}, and then divided by 2 to match the TSE measurements.  Ellipse fitting was not applied in this analysis, as the contrast between the features and the background in the WISPR images was not sufficiently high compared to the TSE images. The measured sizes and radial distances are shown as open squares in Fig. \ref{PlotR}. The best fit slopes were computed individually for each of the five vortex ring sets. In addition, a {\it global} slope was computed by considering all the data as one set, in a manner similar to the TSE measurements,  as shown in Fig. \ref{PlotR}.
The same is applied to the KH waves, which are shown as orange unfilled squares, with a slope of 0.57. The same technique used for the inference of CME sizes in the TSE data was also applied to the five WISPR CMEs, shown as unfilled blue squares, with the same slope of 0.98.

Comparison between the TSE and WISPR data sets in Fig. \ref{PlotR} reveals the following:
(1) For the vortex rings, there is a spatial gap around 3 $R_s$ between the two data sets, with a much shallower slope (0.25) in TSE compared to WISPR (0.81). This is most likely a reflection of their slower speeds as they originate in the low corona and start to accelerate beyond the TSE distance range in which they are detected. (2) The spatial gap for the KH waves between the two data sets is at a similar distance. Here, too, the slope of the TSE KH waves of 0.13 is lower than the 0.57 slope for WISPR, a trend similar to that of the vortex rings. (3) The spatial gap in the CME data is at a larger distance, and the two data sets share the same slope of 0.98, which is steeper than the slopes of vortex rings and KH waves in the distance range over which they were observed.

\subsection{Inference of TSE vortex ring speeds}

Since the evolution of CMEs and KH instabilities starting from the inner corona, is relatively well documented, the speed inferences will be limited to the TSE vortex rings which are the novel features in this work. The distribution of TSE vortex rings across the corona at any given eclipse is a snapshot of their instantaneous dynamic, which unfortunately doesn't yield any speed information. On the other hand the speeds of the WISPR vortex rings, which are apparent plane-of-sky speeds, can be calculated directly from the data, albeit while taking the spacecraft motion into account during the observations. For each WISPR vortex ring presented in Fig. \ref{PlotR}, only a few time stamps were chosen which were selected when the vortex rings were close to the inner edge of the image FoV, thus minimizing projection effects and relative motions. The speeds thus calculated can be viewed as lower-limit estimates as a function of increasing distance over their respective spatial span. For the five WISPR vortex rings' clusters, their speed ($v_2$) is reported in Table \ref{table 1}. Interestingly, their average speed of 249 km/s is almost identical to the lower limit of 250 km/s given by the PSP in-situ solar wind speeds at 10 $R_s$.

\begin{table*}[ht]
\centering
\caption{Linear fit parameters for WISPR and TSE observations. We rewrite Eq. 3 as $\log_{10}(r) = a + b \log_{10}(x)$, where  $r$ is the radial size, $x$ is the distance from the sun limb, $a$ is the intercept and $b$ is the slope for the best line fits for the 5 WISPR clusters and the single TSE slope. The $\pm$ is the standard deviation of the mean. See text for definition of $x_{mid}$.}
\begin{tabular}{|ccccc|c|}
\hline
\textbf{WISPR Clusters} & \textbf{Slope ($b$)} & \textbf{Intercept ($a$)} & \textbf{$x_{\mathrm{mid,2}}$ ($R_s$)} & \textbf{$v_2$ (km s$^{-1}$)} &\textbf{TSE $v_1$ (km s$^{-1}$)}\\
\hline
1 & 1.2694 & $-1.308$  & 13.185  & 320.81 $\pm$ 36.44 & 16.82\\
2 & 2.3276 & $-2.7679$ & 16.550  & 198.82 $\pm$ 20.75 & 11.56 \\
3 & 1.6565 & $-1.5631$ & 6.675   & 274.05 $\pm$ 20.79 & 29.42\\
4 & 0.7936 & $-0.8000$ & 4.550   & 184.39 $\pm$ 6.01 & 23.82\\
5 & 1.1829 & $-1.2176$ & 12.350  & 267.04 $\pm$ 20.15 & 15.35\\
{\bf TSE Average} &&&& & {\bf 19.39 $\pm$ 3.2}\\
{\bf Global} & {\bf 0.8075} & {\bf -0.8417} & {\bf 10.419} & {\bf 249.02 } & {\bf 17.76}\\
\hline
\hline
\textbf{TSE} & \textbf{Slope ($b$)} & \textbf{Intercept ($a$)} & \textbf{$x_{\mathrm{mid,1}}$ ($R_s$)} & &\\
\hline
 & 0.2467 & $-1.1372$ & 0.760 & &\\
\hline
\end{tabular}
\label{table 1}
\end{table*}

Since we assume that the vortex rings in both data sets reflect the same features captured at different distances in the corona, we use mass flux conservation, i.e.

\begin{equation}
    n_1 v_1A_1= n_2v_2A_2
\end{equation}
where $v$ is the speed, $n$ is the number density proportional to $1/V$ where $V$ is the vortex ring's volume, $r$ is its radial size (from Fig 12.), $A$ is its cross-sectional area, and the subscript refers to values from TSE, 1, and WISPR, 2.  Since $V = r \times A$ within some constant factor, depending on the vortex ring geometry, equation (1) then becomes
\begin{equation}
    \frac{v_1}{r_1}= \frac{v_2}{r_2}
\end{equation}

To get the radial sizes, we use the lines of best fit in the log-log plots of Fig. \ref{PlotR}, which can be written as:

\begin{equation}
    r(x) = 10^a x^b
\end{equation}
\noindent
where $a$ and $b$ are obtained from a linear regression of the fitted data points over the distance $x$. For each tracked WISPR vortex ring, we associate their speed $v_2$ with their size at the midpoint distance, which is $r_2(x_{mid,2})$ where $x_{mid} = x_{min} + (x_{max} - x_{min})/2$. As for $r_1$, we take the full set of TSE vortex ring observations from Fig. \ref{PlotR} and use the same definition of midpoint distance ($x_{mid,1}$) as described for the WISPR data to get $r_1(x_{mid,1})$.
Equation (2) then becomes:
\noindent
\begin{equation}
    v_1=v_2 \frac{r_1(x_{mid,1})}{r_2(x_{mid,2})}
\end{equation}
 \noindent

We then use this equation for each of the five different WISPR vortex rings to get five measurements for the TSE speeds, reported in Table \ref{table 1}, which are then  averaged to yield $v_1 = 19.39$ km s$^{-1}$. 
Another approach is to consider all WISPR vortex rings as a single ensemble. We take the average of the five WISPR speeds and use the global best fit of their sizes to get the representative $r_2$ at the global midpoint distance, which then yields $v_1 = 17.76$ km s$^{-1}$. The two $v_1$ values thus inferred are quite close. 
Table 1 shows the parameters from the best-fit lines, with $v_2$  and $x_{mid}$ for each of the five WISPR clusters. The calculated $v_1$ TSE speeds corresponding to those parameters are given in the last column. The lower rows give the slope, intercept and mid point for the TSE vortex data points which were fit by a single line.

\begin{figure}[!h]
\centering
\includegraphics [width=1.0\textwidth]{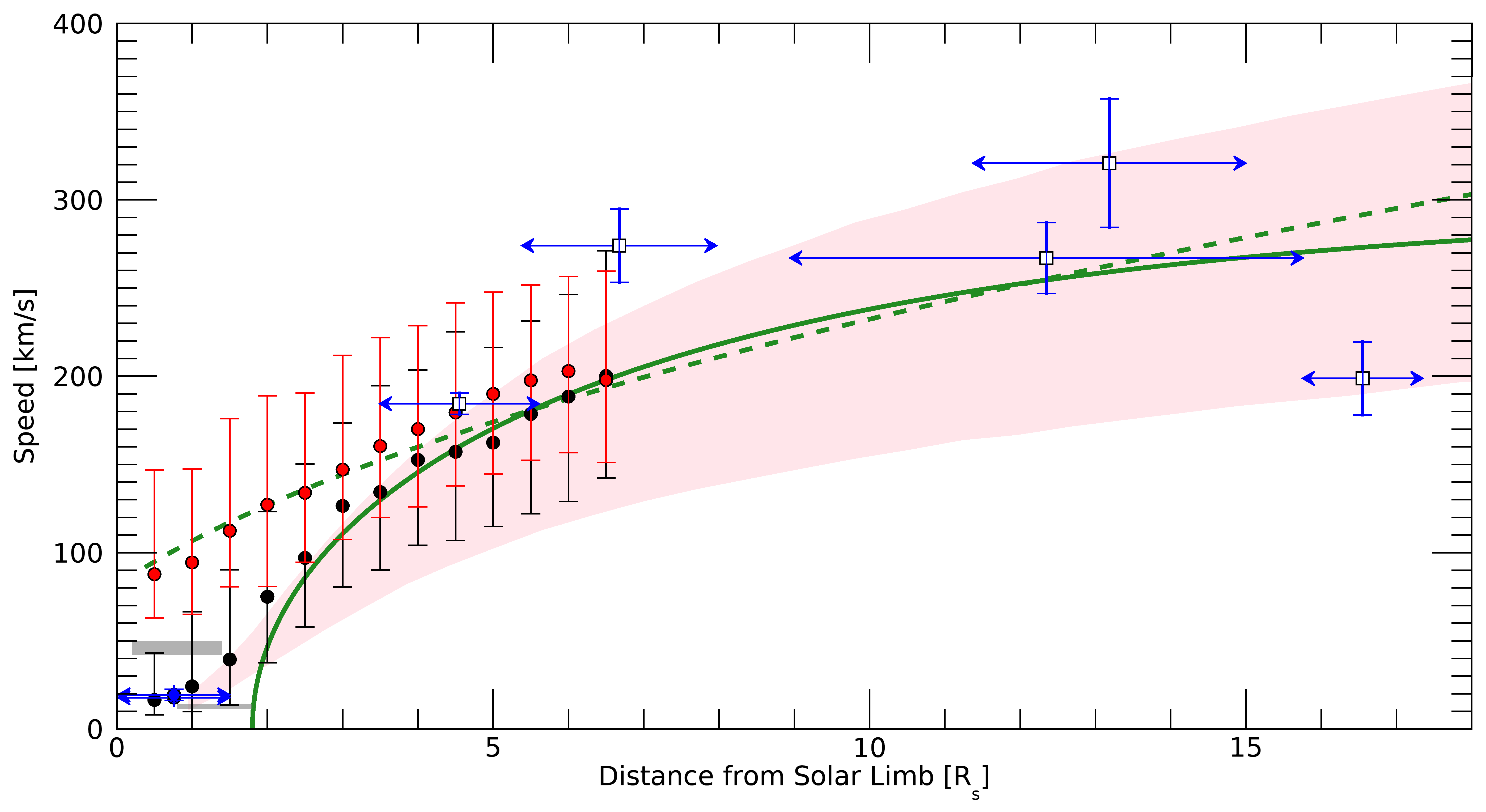}
\caption{\label{naty2}{\footnotesize { 
The average speeds of vortex rings as inferred from Fig. \ref{PlotR} from TSE (filled blue circle) and WISPR (unfilled blue squares) observations in the context of the speed versus distance from the solar limb profile from previous investigations. The blue arrows indicate the distance range spanned by the features used to calculate the average speed. The red and black points with error bars are speeds of outward propagating disturbances identified by \cite{Alzate_2024}. The green curves are speed profiles proposed by \citet{Sheeley_1997} (solid) and used by \citet{Jones_Davila_2009} (dashed). The grey regions are speeds reported by \citet{Seaton_2021} for outflows in off-point EUV observations.  The pink shaded region represents speed estimates based on radio measurements by \citet{Wexler_2020}.
}}}
\end{figure}

As a check of the validity of our approach, the TSE and WISPR vortex ring speeds are plotted within the context of published inferences of solar wind speeds, starting from the very low corona. These are given in Fig. \ref{naty2} as filled blue circles for TSE and as unfilled blue squares for WISPR, with speed and standard deviation along the y-axis and the radial span of the measurements along the x-axis. This figure is a reduced version of the compilation by \cite{Alzate_2024} who used observations from SECCHI/EUVI-COR1-COR2 to track the speed profiles (black and red circles in Figure \ref{naty2}) of propagating brightness disturbances without a specific attribution of their sources or generation mechanism at the Sun.  Their apparent formation occurred in the vicinity of streamers and as low as 0.6 $R_s$ from the solar limb.  Beyond 2 $R_s$ their speed profile became indistinguishable from the speed profile of Sheeley blobs. While our estimates are rather rough, they do indicate that vortex rings propagate at speeds within the ranges given by \cite{Alzate_2024} for the slow solar wind, both in the inner corona and beyond. 

\section{Discussion and Conclusions}

This study was motivated by the recent discoveries from the PSP/WISPR white light images of vortex-like features, KH wave instabilities, and filamentary structures within CMEs. These features were strikingly reminiscent of comparable features discovered earlier in TSE white light images, which were traced back to their origin in the immediate vicinity of prominences in the thermally unstable PCTR.  
Although non-contemporaneous, the uninterrupted spatial span of the FOV of TSE observations, often reaching 10 or more $R_s$ starting from the solar limb,  remains unique. It is currently unrivaled by any other observing platform at present, whether from the ground or space. Luckily, the spatial gap of only 2 -- 3 $R_s$ between the TSE and WISPR data sets was ideally suited for exploring the evolution of vortex rings originating in the low corona away from the Sun. Another fortuitous coincidence is that TSE images are heavily weighted towards the plane of the sky perpendicular to the Earth-Sun observer's line, while WISPR images are acquired while the spacecraft (S/C) is approaching the Sun from a viewpoint perpendicular to the S/C - Sun direction. This geometry enabled orthogonal viewpoints into the exploration of the development and evolution of plasma instabilities in the corona, most notably vortex rings.

There is no doubt that the appearance of vortex rings in the very low corona in the immediate vicinity of prominences, is a consequence of the complex interplay between magnetic fields emerging from the photosphere, into the chromosphere \citep{Tziotziou2023,Murabito2020} and finally interacting with coronal plasmas, as documented in the TSE observations. 

Furthermore, the sharp temperature and density gradients at the prominence-corona transition region (PCTR) create an unstable transition enabling prominence (partial or full) eruptions to release vortex rings, KH instabilities and CMEs. 
Indeed the PCTR has been extensively explored  (see, e.g. \cite{Labrosse_2004, Cirigliano_2004,Terradas_2021}), and the likelihood of plasma instabilities arising in the immediate environment of prominences was reported by \citet{Jun_1995} and \citet{Soler_2012}. They showed how the PCTR is a thermally unstable boundary leading to the development of the Rayleigh-Taylor (RT) instability, which can trigger Kelvin-Helmholtz (KH) wave instabilities and lead to the formation of vortex rings.  \citet{Telloni2019} discussed how vortices can emerge from a Rayleigh-Taylor instability due to ``the onset of secondary KH instabilities".

Invoking KH wave instabilities as the origin of vortex rings is different from magnetic reconnection, which does not lead to vortex rings or a cavity, but rather magnetic bubbles. In fact, analyses of coronagraphic observations and heliospheric imagers have related the intermittent streams of blobs to magnetic reconnection within current sheets in the corona. Yet, this relationship was invoked without being able to observe their sources at the Sun which are invariably hidden behind the occulter. Other interpretations often invoke the tearing mode instability as the underlying mechanism driving the formation of such structures \citep{Furth_1963, Higginson_2018, Reville_2020}. However, clues to additional alternative mechanisms were provided by \citet{Alzate_2021,Alzate_2024}, who extended the coronagraph observations with EUV images to track the initial evolution of features as low down as 0.6 $R_s$ from the solar limb in the proximity of streamers. They also suggested a possible relation to previously proposed mechanisms involving magnetic reconnection events, occurring at the cusp of smaller-scale loops underlying large helmet streamers or between small loops and open-field boundaries low in the corona \citep[e.g.,][]{Antiochos2011, Drake2021}, and/or density variations due to acoustic waves \citep[e.g.,][]{Wexler_2020}.

Also included in this study were comparisons between KH waves and CMEs in white light images from TSE and WISPR observations. Indeed their very close resemblance, in particular the underlying filamentary structures associated of CMEs, lends additional support for the evolution of vortex rings starting from the PCTR.
Indeed, their size distribution indicates that vortex rings, KH waves and CMEs do not disperse as they expand away from the Sun.  The pronounced change in their size evolution near $\approx 1\,R_s$ from the solar limb, as it is increased rapidly as they expanded into the WISPR FOV, also remarked by the slope of the size distribution undergoing a noticeable change, is consistent with the idea that vortex rings and KH waves are entrained with the ambient solar wind and expand practically unscathed, as shown in Fig. \ref{naty2}. This behavior is strikingly reminiscent of the evolution of turbulence amplitude in the low corona, as inferred from remote sensing observations \citep{Telloni_2024} and models \citep{Verdini_Velli_2007,Fox2016}, suggesting the possibility of turbulence driven by PCTR structures in the lower corona.

In conclusion, the fortuitous advent of TSE and WISPR white light images, despite not being co-temporaneous, has offered new insights into the origin and evolution of plasma instabilities, most notably vortex rings, KH waves and CMEs, starting from the PCTR and expanding away from the Sun. These observations have unveiled the underestimated role that the emergence of prominences in the corona play in the origination of turbulence and plasma instabilities in the corona and their expansion into the solar wind.

\pagebreak

AUTHORS' CONTRIBUTIONS:
Everyone on the author list played a valued role in the project.

\begin{acknowledgments}
Support for the TSE observations presented here has been provided over the years by NASA and NSF grants to the Institute for Astronomy of the University of Hawaii, with the latest being NASA grants 80NSSC22K1562 and 80NSSC23K1359, and NSF AGS2313853 for the 2023 and 2024 TSEs. Druckm\"uller from Brno University of Technology was supported by the project ``Innovative Technologies for Smart Low Emission Mobilities", funded as project No. $CZ.02.01.01/00/23$\_$020/0008528$ by Programme Johannes Amos Comenius, call Intersectoral cooperation. This work was also supported by the NASA Parker Solar Probe Program Office for the WISPR program (Contract NNG11EK11I to the U.S. Naval Research Laboratory, NRL). The Parker Solar Probe was designed and built and is now operated by the Johns Hopkins Applied Physics Laboratory as part of NASA’s Living with a Star (LWS) program (contract NNN06AA01C). S.B.S. acknowledges the support from George Mason University (GMU) via the U.S. Naval Research Laboratory contract (N00173-23-2-C603). N.A. acknowledges support from NASA ROSES through HGI grant No. 80NSSC20K1070 and PSP-GI grant No. 80NSSC21K1945.  R.B. was partially supported by NASA grant 80NSSC21K1767.  S.D. acknowledges support from NASA ROSES through PSP-GI grant No. 80NSSC21K1945. We would like to acknowledge the contribution of all the Solar Wind Sherpas whose efforts led to the acquisition of the TSE observations, in particular Dr. Pavel Starha, Mr. Petr Starha, Mr. Matej Starha, Dr. Jana Hoderova, Dr. Ben Boe, and Prof. Adalbert Ding.
\end{acknowledgments}

\bibliography{all_refs.bib}{}

\begin{thebibliography}{}
\expandafter\ifx\csname natexlab\endcsname\relax\def\natexlab#1{#1}\fi
\providecommand{\url}[1]{\href{#1}{#1}}
\providecommand{\dodoi}[1]{doi:~\href{http://doi.org/#1}{\nolinkurl{#1}}}
\providecommand{\doeprint}[1]{\href{http://ascl.net/#1}{\nolinkurl{http://ascl.net/#1}}}
\providecommand{\doarXiv}[1]{\href{https://arxiv.org/abs/#1}{\nolinkurl{https://arxiv.org/abs/#1}}}

\bibitem[{N. {Alzate} {et~al.}(2024){Alzate}, {Di Matteo}, {Morgan}, {Viall}, \& {Vourlidas}}]{Alzate_2024}
{Alzate}, N., {Di Matteo}, S., {Morgan}, H., {Viall}, N., \& {Vourlidas}, A. 2024, \bibinfo{title}{{Connecting the Low to High Corona: Propagating Disturbances as Tracers of the Near-Sun Solar Wind},} \apj, 973, 130, \dodoi{10.3847/1538-4357/ad6601}

\bibitem[{N. {Alzate} {et~al.}(2017){Alzate}, {Habbal}, {Druckm{\"u}ller}, {Emmanouilidis}, \& {Morgan}}]{Alzate_2017}
{Alzate}, N., {Habbal}, S.~R., {Druckm{\"u}ller}, M., {Emmanouilidis}, C., \& {Morgan}, H. 2017, \bibinfo{title}{{Dynamics of Large-scale Coronal Structures as Imaged during the 2012 and 2013 Total Solar Eclipses},} \apj, 848, 84, \dodoi{10.3847/1538-4357/aa8cd2}

\bibitem[{N. {Alzate} {et~al.}(2023){Alzate}, {Morgan}, \& {Di Matteo}}]{Alzate_2023}
{Alzate}, N., {Morgan}, H., \& {Di Matteo}, S. 2023, \bibinfo{title}{{Tracking Nonradial Outflows in Extreme Ultraviolet and White Light Solar Images},} \apj, 945, 116, \dodoi{10.3847/1538-4357/acba08}

\bibitem[{N. {Alzate} {et~al.}(2021){Alzate}, {Morgan}, {Viall}, \& {Vourlidas}}]{Alzate_2021}
{Alzate}, N., {Morgan}, H., {Viall}, N., \& {Vourlidas}, A. 2021, \bibinfo{title}{{Connecting the Low to the High Corona: A Method to Isolate Transients in STEREO/COR1 Images},} \apj, 919, 98, \dodoi{10.3847/1538-4357/ac10ca}

\bibitem[{S.~K. {Antiochos} {et~al.}(2011){Antiochos}, {Miki{\'c}}, {Titov}, {Lionello}, \& {Linker}}]{Antiochos2011}
{Antiochos}, S.~K., {Miki{\'c}}, Z., {Titov}, V.~S., {Lionello}, R., \& {Linker}, J.~A. 2011, \bibinfo{title}{{A Model for the Sources of the Slow Solar Wind},} \apj, 731, 112, \dodoi{10.1088/0004-637X/731/2/112}

\bibitem[{M.~L. {Ascione} {et~al.}(2024){Ascione}, {Gutarra-Leon}, {Shaik}, {Linton}, {Battams}, {Liewer}, \& {Gallagher}}]{Ascione_2024}
{Ascione}, M.~L., {Gutarra-Leon}, A.~J., {Shaik}, S.~B., {et~al.} 2024, \bibinfo{title}{{A Detailed Analysis of a Magnetic Island Observed by WISPR on Parker Solar Probe},} \apj, 973, 12, \dodoi{10.3847/1538-4357/ad5e76}

\bibitem[{T.~E. {Berger} {et~al.}(2012){Berger}, {Liu}, \& {Low}}]{Berger_2012}
{Berger}, T.~E., {Liu}, W., \& {Low}, B.~C. 2012, \bibinfo{title}{{SDO/AIA Detection of Solar Prominence Formation within a Coronal Cavity},} \apjl, 758, L37, \dodoi{10.1088/2041-8205/758/2/L37}

\bibitem[{T.~E. {Berger} {et~al.}(2010){Berger}, {Slater}, {Hurlburt}, {Shine}, {Tarbell}, {Title}, {Lites}, {Okamoto}, {Ichimoto}, {Katsukawa}, {Magara}, {Suematsu}, \& {Shimizu}}]{Berger_2010}
{Berger}, T.~E., {Slater}, G., {Hurlburt}, N., {et~al.} 2010, \bibinfo{title}{{Quiescent Prominence Dynamics Observed with the Hinode Solar Optical Telescope. I. Turbulent Upflow Plumes},} \apj, 716, 1288, \dodoi{10.1088/0004-637X/716/2/1288}

\bibitem[{D.~E. {Billings}(1966){Billings}}]{Billings1966}
{Billings}, D.~E. 1966, {A guide to the solar corona} ({New York: Academic Press})

\bibitem[{G.~E. {Brueckner} {et~al.}(1995){Brueckner}, {Howard}, {Koomen}, {Korendyke}, {Michels}, {Moses}, {Socker}, {Dere}, {Lamy}, {Llebaria}, {Bout}, {Schwenn}, {Simnett}, {Bedford}, \& {Eyles}}]{Brueckner_1995}
{Brueckner}, G.~E., {Howard}, R.~A., {Koomen}, M.~J., {et~al.} 1995, \bibinfo{title}{{The Large Angle Spectroscopic Coronagraph (LASCO)},} \solphys, 162, 357, \dodoi{10.1007/BF00733434}

\bibitem[{G.~M. {Cappello} {et~al.}(2024){Cappello}, {Temmer}, {Vourlidas}, {Braga}, {Liewer}, {Qiu}, {Stenborg}, {Kouloumvakos}, {Veronig}, \& {Bothmer}}]{Cappello_2024}
{Cappello}, G.~M., {Temmer}, M., {Vourlidas}, A., {et~al.} 2024, \bibinfo{title}{{Internal magnetic field structures observed by PSP/WISPR in a filament-related coronal mass ejection},} \aap, 688, A162, \dodoi{10.1051/0004-6361/202449613}

\bibitem[{D. {Cirigliano} {et~al.}(2004){Cirigliano}, {Vial}, \& {Rovira}}]{Cirigliano_2004}
{Cirigliano}, D., {Vial}, J.~C., \& {Rovira}, M. 2004, \bibinfo{title}{{Prominence corona transition region plasma diagnostics from SOHO observations},} \solphys, 223, 95, \dodoi{10.1007/s11207-004-5101-0}

\bibitem[{J.~F. {Drake} {et~al.}(2021){Drake}, {Agapitov}, {Swisdak}, {Badman}, {Bale}, {Horbury}, {Kasper}, {MacDowall}, {Mozer}, {Phan}, {Pulupa}, {Szabo}, \& {Velli}}]{Drake2021}
{Drake}, J.~F., {Agapitov}, O., {Swisdak}, M., {et~al.} 2021, \bibinfo{title}{{Switchbacks as signatures of magnetic flux ropes generated by interchange reconnection in the corona},} \aap, 650, A2, \dodoi{10.1051/0004-6361/202039432}

\bibitem[{M. {Druckm{\"u}ller}(2009){Druckm{\"u}ller}}]{Druckmuller_2009}
{Druckm{\"u}ller}, M. 2009, \bibinfo{title}{{Phase Correlation Method for the Alignment of Total Solar Eclipse Images},} \apj, 706, 1605, \dodoi{10.1088/0004-637X/706/2/1605}

\bibitem[{M. {Druckm{\"u}ller}(2013){Druckm{\"u}ller}}]{Druckmuller_2013}
{Druckm{\"u}ller}, M. 2013, \bibinfo{title}{{A Noise Adaptive Fuzzy Equalization Method for Processing Solar Extreme Ultraviolet Images},} \apjs, 207, 25, \dodoi{10.1088/0067-0049/207/2/25}

\bibitem[{M. {Druckm{\"u}ller} {et~al.}(2014){Druckm{\"u}ller}, {Habbal}, \& {Morgan}}]{Druckmuller_2014}
{Druckm{\"u}ller}, M., {Habbal}, S.~R., \& {Morgan}, H. 2014, \bibinfo{title}{{Discovery of a New Class of Coronal Structures in White Light Eclipse Images},} \apj, 785, 14, \dodoi{10.1088/0004-637X/785/1/14}

\bibitem[{M. {Druckm{\"u}ller} {et~al.}(2006){Druckm{\"u}ller}, {Ru{\v{s}}in}, \& {Minarovjech}}]{Druckmuller_2006}
{Druckm{\"u}ller}, M., {Ru{\v{s}}in}, V., \& {Minarovjech}, M. 2006, \bibinfo{title}{{A new numerical method of total solar eclipse photography processing},} Contributions of the Astronomical Observatory Skalnate Pleso, 36, 131

\bibitem[{C. {Foullon} {et~al.}(2011){Foullon}, {Verwichte}, {Nakariakov}, {Nykyri}, \& {Farrugia}}]{Foullon_2011}
{Foullon}, C., {Verwichte}, E., {Nakariakov}, V.~M., {Nykyri}, K., \& {Farrugia}, C.~J. 2011, \bibinfo{title}{{Magnetic Kelvin-Helmholtz Instability at the Sun},} \apjl, 729, L8, \dodoi{10.1088/2041-8205/729/1/L8}

\bibitem[{N.~J. {Fox} {et~al.}(2016){Fox}, {Velli}, {Bale}, {Decker}, {Driesman}, {Howard}, {Kasper}, {Kinnison}, {Kusterer}, {Lario}, {Lockwood}, {McComas}, {Raouafi}, \& {Szabo}}]{Fox2016}
{Fox}, N.~J., {Velli}, M.~C., {Bale}, S.~D., {et~al.} 2016, \bibinfo{title}{{The Solar Probe Plus Mission: Humanity's First Visit to Our Star},} \ssr, 204, 7, \dodoi{10.1007/s11214-015-0211-6}

\bibitem[{H.~P. {Furth} {et~al.}(1963){Furth}, {Killeen}, \& {Rosenbluth}}]{Furth_1963}
{Furth}, H.~P., {Killeen}, J., \& {Rosenbluth}, M.~N. 1963, \bibinfo{title}{{Finite-Resistivity Instabilities of a Sheet Pinch},} Physics of Fluids, 6, 459, \dodoi{10.1063/1.1706761}

\bibitem[{W. {Grotrian}(1934){Grotrian}}]{Grotrian1934}
{Grotrian}, W. 1934, \bibinfo{title}{{{\"U}ber das Fraunhofersche Spektrum der Sonnenkorona. Mit 10 Abbildungen.},} \zap, 8, 124

\bibitem[{S.~R. {Habbal} {et~al.}(2014){Habbal}, {Morgan}, \& {Druckm{\"u}ller}}]{Habbal_2014}
{Habbal}, S.~R., {Morgan}, H., \& {Druckm{\"u}ller}, M. 2014, \bibinfo{title}{{Exploring the Prominence-Corona Connection and its Expansion into the Outer Corona Using Total Solar Eclipse Observations},} \apj, 793, 119, \dodoi{10.1088/0004-637X/793/2/119}

\bibitem[{P. {Hess} {et~al.}(2020){Hess}, {Rouillard}, {Kouloumvakos}, {Liewer}, {Zhang}, {Dhakal}, {Stenborg}, {Colaninno}, \& {Howard}}]{Hess_2020}
{Hess}, P., {Rouillard}, A.~P., {Kouloumvakos}, A., {et~al.} 2020, \bibinfo{title}{{WISPR Imaging of a Pristine CME},} \apjs, 246, 25, \dodoi{10.3847/1538-4365/ab4ff0}

\bibitem[{A.~K. {Higginson} \& B.~J. {Lynch}(2018){Higginson} \& {Lynch}}]{Higginson_2018}
{Higginson}, A.~K., \& {Lynch}, B.~J. 2018, \bibinfo{title}{{Structured Slow Solar Wind Variability: Streamer-blob Flux Ropes and Torsional Alfv{\'e}n Waves},} \apj, 859, 6, \dodoi{10.3847/1538-4357/aabc08}

\bibitem[{R.~A. {Howard} {et~al.}(2022){Howard}, {Stenborg}, {Vourlidas}, {Gallagher}, {Linton}, {Hess}, {Rich}, \& {Liewer}}]{Howard_2022}
{Howard}, R.~A., {Stenborg}, G., {Vourlidas}, A., {et~al.} 2022, \bibinfo{title}{{Overview of the Remote Sensing Observations from PSP Solar Encounter 10 with Perihelion at 13.3 R$_{{\ensuremath{\odot}}}$},} \apj, 936, 43, \dodoi{10.3847/1538-4357/ac7ff5}

\bibitem[{R.~A. {Howard} {et~al.}(2019){Howard}, {Vourlidas}, {Bothmer}, {Colaninno}, {DeForest}, {Gallagher}, {Hall}, {Hess}, {Higginson}, {Korendyke}, {Kouloumvakos}, {Lamy}, {Liewer}, {Linker}, {Linton}, {Penteado}, {Plunkett}, {Poirier}, {Raouafi}, {Rich}, {Rochus}, {Rouillard}, {Socker}, {Stenborg}, {Thernisien}, \& {Viall}}]{howard_2019}
{Howard}, R.~A., {Vourlidas}, A., {Bothmer}, V., {et~al.} 2019, \bibinfo{title}{{Near-Sun observations of an F-corona decrease and K-corona fine structure},} \nat, 576, 232, \dodoi{10.1038/s41586-019-1807-x}

\bibitem[{R.~M.~E. {Illing} \& A.~J. {Hundhausen}(1985){Illing} \& {Hundhausen}}]{IllingH_1985}
{Illing}, R.~M.~E., \& {Hundhausen}, A.~J. 1985, \bibinfo{title}{{Observation of a coronal transient from 1.2 to 6 solar radii},} \jgr, 90, 275, \dodoi{10.1029/JA090iA01p00275}

\bibitem[{M. {Janssen}(1869{\natexlab{a}}){Janssen}}]{janssen1869a}
{Janssen}, M. 1869{\natexlab{a}}, \bibinfo{title}{{The Total Solar Eclipse of August 1868. Part I.},} Astronomical register, 7, 107

\bibitem[{M. {Janssen}(1869{\natexlab{b}}){Janssen}}]{janssen1869b}
{Janssen}, M. 1869{\natexlab{b}}, \bibinfo{title}{{The Total Solar Eclipse of August 1868. Part II.},} Astronomical register, 7, 131

\bibitem[{S.~I. {Jones} \& J.~M. {Davila}(2009){Jones} \& {Davila}}]{Jones_Davila_2009}
{Jones}, S.~I., \& {Davila}, J.~M. 2009, \bibinfo{title}{{Localized Plasma Density Enhancements Observed in STEREO COR1},} \apj, 701, 1906, \dodoi{10.1088/0004-637X/701/2/1906}

\bibitem[{B.-I. {Jun} {et~al.}(1995){Jun}, {Norman}, \& {Stone}}]{Jun_1995}
{Jun}, B.-I., {Norman}, M.~L., \& {Stone}, J.~M. 1995, \bibinfo{title}{{A Numerical Study of Rayleigh-Taylor Instability in Magnetic Fluids},} \apj, 453, 332, \dodoi{10.1086/176393}

\bibitem[{H. {Kimura} \& I. {Mann}(1998){Kimura} \& {Mann}}]{Kimura_1998}
{Kimura}, H., \& {Mann}, I. 1998, \bibinfo{title}{{Brightness of the solar F-corona},} Earth, Planets and Space, 50, 493, \dodoi{10.1186/BF03352140}

\bibitem[{N. {Labrosse} \& P. {Gouttebroze}(2004){Labrosse} \& {Gouttebroze}}]{Labrosse_2004}
{Labrosse}, N., \& {Gouttebroze}, P. 2004, \bibinfo{title}{{Non-LTE Radiative Transfer in Model Prominences. I. Integrated Intensities of He I Triplet Lines},} \apj, 617, 614, \dodoi{10.1086/425168}

\bibitem[{J.~R. {Lemen} {et~al.}(2012){Lemen}, {Title}, {Akin}, {Boerner}, {Chou}, {Drake}, {Duncan}, {Edwards}, {Friedlaender}, {Heyman}, {Hurlburt}, {Katz}, {Kushner}, {Levay}, {Lindgren}, {Mathur}, {McFeaters}, {Mitchell}, {Rehse}, {Schrijver}, {Springer}, {Stern}, {Tarbell}, {Wuelser}, {Wolfson}, {Yanari}, {Bookbinder}, {Cheimets}, {Caldwell}, {Deluca}, {Gates}, {Golub}, {Park}, {Podgorski}, {Bush}, {Scherrer}, {Gummin}, {Smith}, {Auker}, {Jerram}, {Pool}, {Soufli}, {Windt}, {Beardsley}, {Clapp}, {Lang}, \& {Waltham}}]{Lemen2012}
{Lemen}, J.~R., {Title}, A.~M., {Akin}, D.~J., {et~al.} 2012, \bibinfo{title}{{The Atmospheric Imaging Assembly (AIA) on the Solar Dynamics Observatory (SDO)},} \solphys, 275, 17, \dodoi{10.1007/s11207-011-9776-8}

\bibitem[{X. {Li} {et~al.}(2012){Li}, {Morgan}, {Leonard}, \& {Jeska}}]{Li_2012}
{Li}, X., {Morgan}, H., {Leonard}, D., \& {Jeska}, L. 2012, \bibinfo{title}{{A Solar Tornado Observed by AIA/SDO: Rotational Flow and Evolution of Magnetic Helicity in a Prominence and Cavity},} \apjl, 752, L22, \dodoi{10.1088/2041-8205/752/2/L22}

\bibitem[{P.~C. {Liewer} {et~al.}(2024){Liewer}, {Gallagher}, {Stenborg}, {Linton}, {Qiu}, {Vourlidas}, {Ascione}, \& {Velli}}]{Liewer_2024}
{Liewer}, P.~C., {Gallagher}, B.~M., {Stenborg}, G., {et~al.} 2024, \bibinfo{title}{{Evidence of Continuous Reconnection along a Helmet Streamer Current Sheet Observed by WISPR on Parker Solar Probe},} \apj, 970, 79, \dodoi{10.3847/1538-4357/ad509b}

\bibitem[{P.~C. {Liewer} {et~al.}(2023){Liewer}, {Vourlidas}, {Stenborg}, {Howard}, {Qiu}, {Penteado}, {Panasenco}, \& {Braga}}]{Liewer_2023}
{Liewer}, P.~C., {Vourlidas}, A., {Stenborg}, G., {et~al.} 2023, \bibinfo{title}{{Structure of the Plasma near the Heliospheric Current Sheet as Seen by WISPR/Parker Solar Probe from inside the Streamer Belt},} \apj, 948, 24, \dodoi{10.3847/1538-4357/acc8c7}

\bibitem[{M. {Liggett} \& H. {Zirin}(1984){Liggett} \& {Zirin}}]{Liggett_Zirin_1984}
{Liggett}, M., \& {Zirin}, H. 1984, \bibinfo{title}{{Rotation in Prominences},} \solphys, 91, 259, \dodoi{10.1007/BF00146298}

\bibitem[{M. {Murabito} {et~al.}(2020){Murabito}, {Shetye}, {Stangalini}, {Verwichte}, {Arber}, {Ermolli}, {Giorgi}, \& {Goffrey}}]{Murabito2020}
{Murabito}, M., {Shetye}, J., {Stangalini}, M., {et~al.} 2020, \bibinfo{title}{{Unveiling the magnetic nature of chromospheric vortices},} \aap, 639, A59, \dodoi{10.1051/0004-6361/202038360}

\bibitem[{G. {Nistic{\`o}} {et~al.}(2020){Nistic{\`o}}, {Bothmer}, {Vourlidas}, {Liewer}, {Thernisien}, {Stenborg}, \& {Howard}}]{Nistico_2020}
{Nistic{\`o}}, G., {Bothmer}, V., {Vourlidas}, A., {et~al.} 2020, \bibinfo{title}{{Simulating White-Light Images of Coronal Structures for Parker Solar Probe/WISPR: Study of the Total Brightness Profiles},} \solphys, 295, 63, \dodoi{10.1007/s11207-020-01626-y}

\bibitem[{L. {Ofman} \& B.~J. {Thompson}(2011){Ofman} \& {Thompson}}]{Ofman_2011}
{Ofman}, L., \& {Thompson}, B.~J. 2011, \bibinfo{title}{{SDO/AIA Observation of Kelvin-Helmholtz Instability in the Solar Corona},} \apjl, 734, L11, \dodoi{10.1088/2041-8205/734/1/L11}

\bibitem[{Y. {{\"O}hman}(1968){{\"O}hman}}]{Ohman_1968}
{{\"O}hman}, Y. 1968, \bibinfo{title}{{An H{\ensuremath{\alpha}} Filament observed against the Chromosphere at the Limb},} \solphys, 3, 354, \dodoi{10.1007/BF00155171}

\bibitem[{Y. {{\"O}hman}(1969){{\"O}hman}}]{Ohman_1969}
{{\"O}hman}, Y. 1969, \bibinfo{title}{{Observations of Rotational Motion in Prominences},} \solphys, 9, 427, \dodoi{10.1007/BF02391666}

\bibitem[{E. {Paouris} {et~al.}(2024){Paouris}, {Stenborg}, {Linton}, {Vourlidas}, {Howard}, \& {Raouafi}}]{Paouris_2024}
{Paouris}, E., {Stenborg}, G., {Linton}, M.~G., {et~al.} 2024, \bibinfo{title}{{First Direct Imaging of a Kelvin{\textendash}Helmholtz Instability by PSP/WISPR},} \apj, 964, 139, \dodoi{10.3847/1538-4357/ad2208}

\bibitem[{E. {Pettit} \& S.~B. {Nicholson}(1925){Pettit} \& {Nicholson}}]{Pettit_1925}
{Pettit}, E., \& {Nicholson}, S.~B. 1925, \bibinfo{title}{{Radiation Measurements of the Solar Corona January 24, 1925},} \apj, 62, 202, \dodoi{10.1086/142926}

\bibitem[{V. {R{\'e}ville} {et~al.}(2020){R{\'e}ville}, {Velli}, {Rouillard}, {Lavraud}, {Tenerani}, {Shi}, \& {Strugarek}}]{Reville_2020}
{R{\'e}ville}, V., {Velli}, M., {Rouillard}, A.~P., {et~al.} 2020, \bibinfo{title}{{Tearing Instability and Periodic Density Perturbations in the Slow Solar Wind},} \apjl, 895, L20, \dodoi{10.3847/2041-8213/ab911d}

\bibitem[{M. {Ryutova} {et~al.}(2010){Ryutova}, {Berger}, {Frank}, {Tarbell}, \& {Title}}]{Ryutova_2010}
{Ryutova}, M., {Berger}, T., {Frank}, Z., {Tarbell}, T., \& {Title}, A. 2010, \bibinfo{title}{{Observation of Plasma Instabilities in Quiescent Prominences},} \solphys, 267, 75, \dodoi{10.1007/s11207-010-9638-9}

\bibitem[{D. {Schmidt} {et~al.}(2025){Schmidt}, {Schad}, {Yurchyshyn}, {Gorceix}, {Rimmele}, \& {Goode}}]{Schmidt_2025}
{Schmidt}, D., {Schad}, T.~A., {Yurchyshyn}, V., {et~al.} 2025, \bibinfo{title}{{Observations of fine coronal structures with high-order solar adaptive optics},} Nature Astronomy, \dodoi{10.1038/s41550-025-02564-0}

\bibitem[{D.~B. {Seaton} {et~al.}(2021){Seaton}, {Hughes}, {Tadikonda}, {Caspi}, {DeForest}, {Krimchansky}, {Hurlburt}, {Seguin}, \& {Slater}}]{Seaton_2021}
{Seaton}, D.~B., {Hughes}, J.~M., {Tadikonda}, S.~K., {et~al.} 2021, \bibinfo{title}{{The Sun's dynamic extended corona observed in extreme ultraviolet},} Nature Astronomy, 5, 1029, \dodoi{10.1038/s41550-021-01427-8}

\bibitem[{S.~B. {Shaik} {et~al.}(2024){Shaik}, {Linton}, {Gibson}, {Hess}, {Colaninno}, {Stenborg}, {Braga}, \& {Palmerio}}]{Shaik_2024}
{Shaik}, S.~B., {Linton}, M.~G., {Gibson}, S.~E., {et~al.} 2024, \bibinfo{title}{{A Study on the Nested Rings CME Structure Observed by the WISPR Imager Onboard Parker Solar Probe},} \apj, 976, 179, \dodoi{10.3847/1538-4357/ad8354}

\bibitem[{J. {Sheeley} \& Y.~M. {Wang}(2002){Sheeley} \& {Wang}}]{2002Sheeley&Wang}
{Sheeley}, N.~R., J., \& {Wang}, Y.~M. 2002, \bibinfo{title}{{Characteristics of Coronal Inflows},} \apj, 579, 874, \dodoi{10.1086/342923}

\bibitem[{J. {Sheeley} \& Y.~M. {Wang}(2007){Sheeley} \& {Wang}}]{2007Sheeley&Wang}
{Sheeley}, N.~R., J., \& {Wang}, Y.~M. 2007, \bibinfo{title}{{In/Out Pairs and the Detachment of Coronal Streamers},} \apj, 655, 1142, \dodoi{10.1086/510323}

\bibitem[{N.~R. {Sheeley} {et~al.}(1997{\natexlab{a}}){Sheeley}, {Wang}, {Hawley}, {Brueckner}, {Dere}, {Howard}, {Koomen}, {Korendyke}, {Michels}, {Paswaters}, {Socker}, {St. Cyr}, {Wang}, {Lamy}, {Llebaria}, {Schwenn}, {Simnett}, {Plunkett}, \& {Biesecker}}]{1997Sheeley}
{Sheeley}, N.~R., {Wang}, Y.~M., {Hawley}, S.~H., {et~al.} 1997{\natexlab{a}}, \bibinfo{title}{{Measurements of Flow Speeds in the Corona Between 2 and 30 R$_{{\ensuremath{\odot}}}$},} \apj, 484, 472, \dodoi{10.1086/304338}

\bibitem[{N.~R. {Sheeley} {et~al.}(1997{\natexlab{b}}){Sheeley}, {Wang}, {Hawley}, {Brueckner}, {Dere}, {Howard}, {Koomen}, {Korendyke}, {Michels}, {Paswaters}, {Socker}, {St. Cyr}, {Wang}, {Lamy}, {Llebaria}, {Schwenn}, {Simnett}, {Plunkett}, \& {Biesecker}}]{Sheeley_1997}
{Sheeley}, N.~R., {Wang}, Y.~M., {Hawley}, S.~H., {et~al.} 1997{\natexlab{b}}, \bibinfo{title}{{Measurements of Flow Speeds in the Corona Between 2 and 30 R$_{{\ensuremath{\odot}}}$},} \apj, 484, 472, \dodoi{10.1086/304338}

\bibitem[{N.~R. {Sheeley} {et~al.}(2009){Sheeley}, {Lee}, {Casto}, {Wang}, \& {Rich}}]{Sheeley_2009}
{Sheeley}, Jr., N.~R., {Lee}, D.~D.~H., {Casto}, K.~P., {Wang}, Y.~M., \& {Rich}, N.~B. 2009, \bibinfo{title}{{The Structure of Streamer Blobs},} \apj, 694, 1471, \dodoi{10.1088/0004-637X/694/2/1471}

\bibitem[{R. {Soler} {et~al.}(2012){Soler}, {Ballester}, \& {Parenti}}]{Soler_2012}
{Soler}, R., {Ballester}, J.~L., \& {Parenti}, S. 2012, \bibinfo{title}{{Stability of thermal modes in cool prominence plasmas},} \aap, 540, A7, \dodoi{10.1051/0004-6361/201118492}

\bibitem[{E. {Tandberg-Hanssen}(1995){Tandberg-Hanssen}}]{Tandberg-Hanssen_1995}
{Tandberg-Hanssen}, E. 1995, {The nature of solar prominences}, Vol. 199, \dodoi{10.1007/978-94-017-3396-0}

\bibitem[{D. {Telloni} {et~al.}(2019){Telloni}, {Carbone}, {Bemporad}, \& {Antonucci}}]{Telloni2019}
{Telloni}, D., {Carbone}, F., {Bemporad}, A., \& {Antonucci}, E. 2019, \bibinfo{title}{{Evidence for Rayleigh-Taylor Plasma Instability at the Front of Solar Coronal Mass Ejections},} Atmosphere, 10, 468, \dodoi{10.3390/atmos10080468}

\bibitem[{D. {Telloni} {et~al.}(2024){Telloni}, {Sorriso-Valvo}, {Zank}, {Velli}, {Andretta}, {Perrone}, {Marino}, {Carbone}, {Vecchio}, {Adhikari}, {Zhao}, {Guastavino}, {Camattari}, {Shi}, {Sioulas}, {Huang}, {Romoli}, {Antonucci}, {Da Deppo}, {Fineschi}, {Grimani}, {Heinzel}, {Moses}, {Naletto}, {Nicolini}, {Spadaro}, {Stangalini}, {Teriaca}, {Uslenghi}, {Abbo}, {Auch{\`e}re}, {Cuadrado}, {Berlicki}, {Bruno}, {Burtovoi}, {Capobianco}, {Casini}, {Casti}, {Chioetto}, {Corso}, {D'Amicis}, {De Leo}, {Fabi}, {Frassati}, {Frassetto}, {Giordano}, {Guglielmino}, {Jerse}, {Landini}, {Liberatore}, {Magli}, {Massone}, {Nistic{\`o}}, {Pancrazzi}, {Pelizzo}, {Peter}, {Plainaki}, {Poletto}, {Reale}, {Romano}, {Russano}, {Sasso}, {Sch{\"u}hle}, {Solanki}, {Strachan}, {Straus}, {Susino}, {Ventura}, {Volpicelli}, {Woch}, {Zangrilli}, {Zimbardo}, \& {Zuppella}}]{Telloni_2024}
{Telloni}, D., {Sorriso-Valvo}, L., {Zank}, G.~P., {et~al.} 2024, \bibinfo{title}{{Metis Observation of the Onset of Fully Developed Turbulence in the Solar Corona},} \apjl, 973, L48, \dodoi{10.3847/2041-8213/ad5a8c}

\bibitem[{J. {Terradas} {et~al.}(2021){Terradas}, {Luna}, {Soler}, {Oliver}, {Carbonell}, \& {Ballester}}]{Terradas_2021}
{Terradas}, J., {Luna}, M., {Soler}, R., {et~al.} 2021, \bibinfo{title}{{One-dimensional prominence threads. I. Equilibrium models},} \aap, 653, A95, \dodoi{10.1051/0004-6361/202039905}

\bibitem[{K. {Tziotziou} {et~al.}(2023){Tziotziou}, {Scullion}, {Shelyag}, {Steiner}, {Khomenko}, {Tsiropoula}, {Canivete Cuissa}, {Wedemeyer}, {Kontogiannis}, {Yadav}, {Kitiashvili}, {Skirvin}, {Dakanalis}, {Kosovichev}, \& {Fedun}}]{Tziotziou2023}
{Tziotziou}, K., {Scullion}, E., {Shelyag}, S., {et~al.} 2023, \bibinfo{title}{{Vortex Motions in the Solar Atmosphere},} \ssr, 219, 1, \dodoi{10.1007/s11214-022-00946-8}

\bibitem[{A. {Verdini} \& M. {Velli}(2007){Verdini} \& {Velli}}]{Verdini_Velli_2007}
{Verdini}, A., \& {Velli}, M. 2007, \bibinfo{title}{{Alfv{\'e}n Waves and Turbulence in the Solar Atmosphere and Solar Wind},} \apj, 662, 669, \dodoi{10.1086/510710}

\bibitem[{A. {Vourlidas} \& R.~A. {Howard}(2006){Vourlidas} \& {Howard}}]{Vourlidas_2006}
{Vourlidas}, A., \& {Howard}, R.~A. 2006, \bibinfo{title}{{The Proper Treatment of Coronal Mass Ejection Brightness: A New Methodology and Implications for Observations},} \apj, 642, 1216, \dodoi{10.1086/501122}

\bibitem[{A. {Vourlidas} {et~al.}(2013){Vourlidas}, {Lynch}, {Howard}, \& {Li}}]{Vourlidas_2013}
{Vourlidas}, A., {Lynch}, B.~J., {Howard}, R.~A., \& {Li}, Y. 2013, \bibinfo{title}{{How Many CMEs Have Flux Ropes? Deciphering the Signatures of Shocks, Flux Ropes, and Prominences in Coronagraph Observations of CMEs},} \solphys, 284, 179, \dodoi{10.1007/s11207-012-0084-8}

\bibitem[{A. {Vourlidas} {et~al.}(2016){Vourlidas}, {Howard}, {Plunkett}, {Korendyke}, {Thernisien}, {Wang}, {Rich}, {Carter}, {Chua}, {Socker}, {Linton}, {Morrill}, {Lynch}, {Thurn}, {Van Duyne}, {Hagood}, {Clifford}, {Grey}, {Velli}, {Liewer}, {Hall}, {DeJong}, {Mikic}, {Rochus}, {Mazy}, {Bothmer}, \& {Rodmann}}]{Vourlidas2016}
{Vourlidas}, A., {Howard}, R.~A., {Plunkett}, S.~P., {et~al.} 2016, \bibinfo{title}{{The Wide-Field Imager for Solar Probe Plus (WISPR)},} \ssr, 204, 83, \dodoi{10.1007/s11214-014-0114-y}

\bibitem[{D. {Wexler} {et~al.}(2020){Wexler}, {Imamura}, {Efimov}, {Song}, {Lukanina}, {Ando}, {Jensen}, {Vierinen}, \& {Coster}}]{Wexler_2020}
{Wexler}, D., {Imamura}, T., {Efimov}, A., {et~al.} 2020, \bibinfo{title}{{Coronal Electron Density Fluctuations Inferred from Akatsuki Spacecraft Radio Observations},} \solphys, 295, 111, \dodoi{10.1007/s11207-020-01677-1}

\bibitem[{B.~E. {Wood} {et~al.}(2023){Wood}, {Hess}, {Chen}, \& {Hu}}]{Wood_2023}
{Wood}, B.~E., {Hess}, P., {Chen}, Y., \& {Hu}, Q. 2023, \bibinfo{title}{{Sequential Small Coronal Mass Ejections Observed In Situ and in White-Light Images by Parker Solar Probe},} \apj, 953, 123, \dodoi{10.3847/1538-4357/ace259}

\end{thebibliography}
\bibliographystyle{aasjournalv7}



\end{document}